\documentclass[a4paper,fleqn,usenatbib]{mnras}

\usepackage[T1]{fontenc}
\usepackage{ae,aecompl}

\usepackage{graphicx}	
\usepackage{amsmath}	
\usepackage{amssymb}	
\usepackage{longtable}
\usepackage[normalem]{ulem}

\title[Dynamical analysis of  A3407 + A3408]{Dynamical analysis of the cluster pair: A3407 + A3408 }
\author[R.S. Nascimento et al.] {R.S. Nascimento$^{1}$\thanks{E-mail: rnascimento@astro.ufrj.br}, Ribeiro, A.L.B.$^{2}$,
M. Trevisan$^{3,4}$, E. R. Carrasco$^{5}$, \and
H. Plana$^{2}$, and R. Dupke$^{6,7,8}$ \\\\
$^{1}$ Observat\'orio do Valongo, Universidade Federal do Rio de Janeiro, Rio de Janeiro-RJ, Brasil\\
$^{2}$ Laborat\'orio de Astrof\'{\i}sica Te\'orica e Observacional, Universidade Estadual de Santa Cruz -- 45650-000, Ilh\'eus-BA, Brasil\\
$^{3}$ Instituto Nacional de Pesquisas Espaciais, S\~ao Jos\'e dos Campos-SP, Brasil\\
$^{4}$ Institut d'Astrophysique de Paris (UMR 7095: CNRS \& UPMC), 98 bis Bd Arago, 75014 Paris, France\\
$^{5}$ Gemini Observatory, Southern Operations Center, AURA, Casilla 603, La Serena, Chile\\
$^{6}$ Observat\'orio Nacional, CP 20921-400, Rio de Janeiro-RJ, Brasil\\
$^{7}$ Dept. of Astronomy, University of Michigan, 500 Church St., Ann Arbor, MI 48109, USA\\
$^{8}$ Eureka Scientific Inc., 2452 Delmer St. Suite 100, Oakland, CA 94602, USA}

\date{Accepted XXX. Received YYY; in original form ZZZ}

\pubyear{2016}

\begin{document}
\label{firstpage}
\pagerange{\pageref{firstpage}--\pageref{lastpage}}
\maketitle
\begin{abstract}

We carried out a dynamical study of the galaxy cluster pair A3407 \& A3408 based on a spectroscopic survey obtained
with the 4 meter Blanco telescope at the CTIO, plus 6dF data, and ROSAT All-Sky-Survey. The sample consists of 122 member galaxies brighter than $m_R=20$. Our main goal is to probe the galaxy dynamics in this field and verify if the sample constitutes a single galaxy system or corresponds to an ongoing merging process. Statistical tests were applied to clusters members showing that both the composite system A3407 + A3408 as well as each individual cluster have Gaussian velocity distribution. A velocity gradient of $\sim 847\pm 114$ $\rm km\;s^{-1}$ was identified around the principal axis of the projected distribution of galaxies, indicating that the global field may be rotating. Applying the KMM algorithm to the distribution of galaxies we found that the solution with two clusters is better than the single unit solution at the 99\% c.l. This is consistent with the X-ray distribution around this field, which shows no common X-ray halo involving A3407 and A3408. We also estimated virial masses and applied a two-body model to probe the dynamics of the pair. The more likely scenario is that in which the pair is gravitationally bound and probably experiences a collapse phase, with the cluster cores crossing in less than $\sim$1 $h^{-1}$ Gyr, a pre-merger scenario. The complex X-ray morphology, the gas temperature, and some signs of galaxy evolution in A3408  suggests a post-merger scenario, with cores having crossed each other $\sim 1.65 h^{-1}$Gyr ago, as an alternative solution.

\end{abstract}


\section{Introduction}

Clusters of galaxies are good tracers of the large-scale distribution of matter. They are the largest gravitationally bound systems in the universe, constraining both structure formation and the composition of the universe \citep[e.g.][]{V05,A11,KB}. These systems also constitute important environments for the study of galaxy formation and evolution. In the hierarchical scenario, clusters are relatively recent structures collapsing at $z \lesssim 2$ \citep[e.g.][]{CW}, and growing at the intersections of cosmic filaments \citep[e.g.][]{SWJ,AM9}. In the ${\rm \Lambda}$CDM scenario, structures form in a bottom-up fashion: more massive galaxy systems assemble their mass from the merging of less massive ones \citep[e.g.][]{DSW,NJO,CMW}. Con\-ti\-nuous galaxy interaction for period longer than the re\-la\-xa\-tion time tend to distribute the velocities of the galaxy members towards a Gaussian distribution \citep[e.g.][]{BB}. This provides a way to access the dynamical state of galaxy clusters by studying their velocity distributions. Usually, distributions are well approximated by a Gaussian in the virialized (more central) regions of clusters \citep[e.g.][]{Yah77}, while in peripheral areas they can show deviations from Gaussianity \citep[e.g.][]{RLP}. This indicates that the central parts are probably in dynamical equilibrium, when outskirts continue to accumulate matter from the surroundings. This accretion of matter, in the form of galaxies or groups of galaxies from the neighbourhood, seems to occur along giant filamentary structures \citep[e.g.][]{KRL}. This suggests that the formation of a galaxy cluster is a continuous process that takes place through mergers and encounters in greater or lesser proportions. Some outstanding e\-xam\-ples of this are the so-called ''Bullet Cluster'' (1E 0657-56) \citep{CBG, Jee07}, and other clusters like Cl 0152-1357, MS 1054 \citep{Jee05a, Jee05b}, and Abell 520 \citep{MA05,MA07}.

To understand the impact of mergers on cluster evolution it is important to study the process at different epochs. In the literature, only a few early merging clusters have been found up to now \citep{KA15}. Examples of this are the pairs Abell 222-223 \citep[e.g.][]{WE08}, and Abell 399-401 \citep[e.g.][]{FKT,FTH}. Systems like Abell 3407-3408 (hereafter A3407-A3408), relatively isolated in the field, may provide an invaluable opportunity to study early signatures of merging clusters. This pair lies in a largely unexplored low galactic latitude section ($b \approx 17.57^\circ$) of the southern sky, where just few optical and X-ray observations have been performed. All available information may be summarised as follows: (i) The morphological classification of A3407 and A3408 is Bautz-Morgan type I, and type I-II, respectively \citep{Abe89}, suggesting they are relaxed to moderately relaxed systems. (ii) On the other hand, the study of \citet{GCF} suggests that A3407 and A3408 are interacting and may form a single system. (iii) \citet{CH} discovered an arc-like feature (z = 0.073) near to the center of A3408. This result was further confirmed by \citet{CKH} and \citet{CSC}. (iv)  The pair has been detected by the Rosat-All-Sky-Survey \citep{EVB} and A3408 by ASCA \citep{KHH}. (v) Finally, the estimated mass of A3408, evaluated from ASCA X-ray observations and enclosed within the arc radius, represents 18\%--45\% of the dynamical mass computed by \citet{CKH}. In their estimation, \citet{CKH} assume that the center of the cluster potential coincides with the Brightest cluster galaxy (BCG), when this one is $\sim$60$^{\prime\prime}$ off the X-ray center \citep{KHH}.  
 
In this work, we present new radial velocities for galaxies around the galaxy cluster pair A3407 \& A3408. Our main goal is to probe the galaxy dynamics in this field and verify if the sample constitutes a single galaxy system or corresponds to an ongoing merging process, improving the understanding of this system. The paper is organized as follows: in Section 2, we present the observations, data reduction  and the me\-tho\-do\-lo\-gy used to find the galaxy redshifts; in Section 3, we present a study of the velocity distribution, covering membership determination, normality tests, and identification of significant gaps; in Section 4 we study subclustering in the field; in Section 5 we present a dynamical analysis, covering the virialization properties and the two-body model applied to the pair A3407-A3408; and in Section 6, we discuss our results.\footnote{Throughout this paper we assume a $\Lambda$CDM cosmology whith the cosmological parameters $\Omega_M=0.3$, $\Omega_\Lambda=0.7$ and ${\rm H_0=100\;h}$ ${\rm km\;s^{-1}\;Mpc^{-1}}$.}

\section{Observations and Data Reduction}

\subsection{Observations}
 
\begin{figure*}
\centering
 \includegraphics[width=160mm]{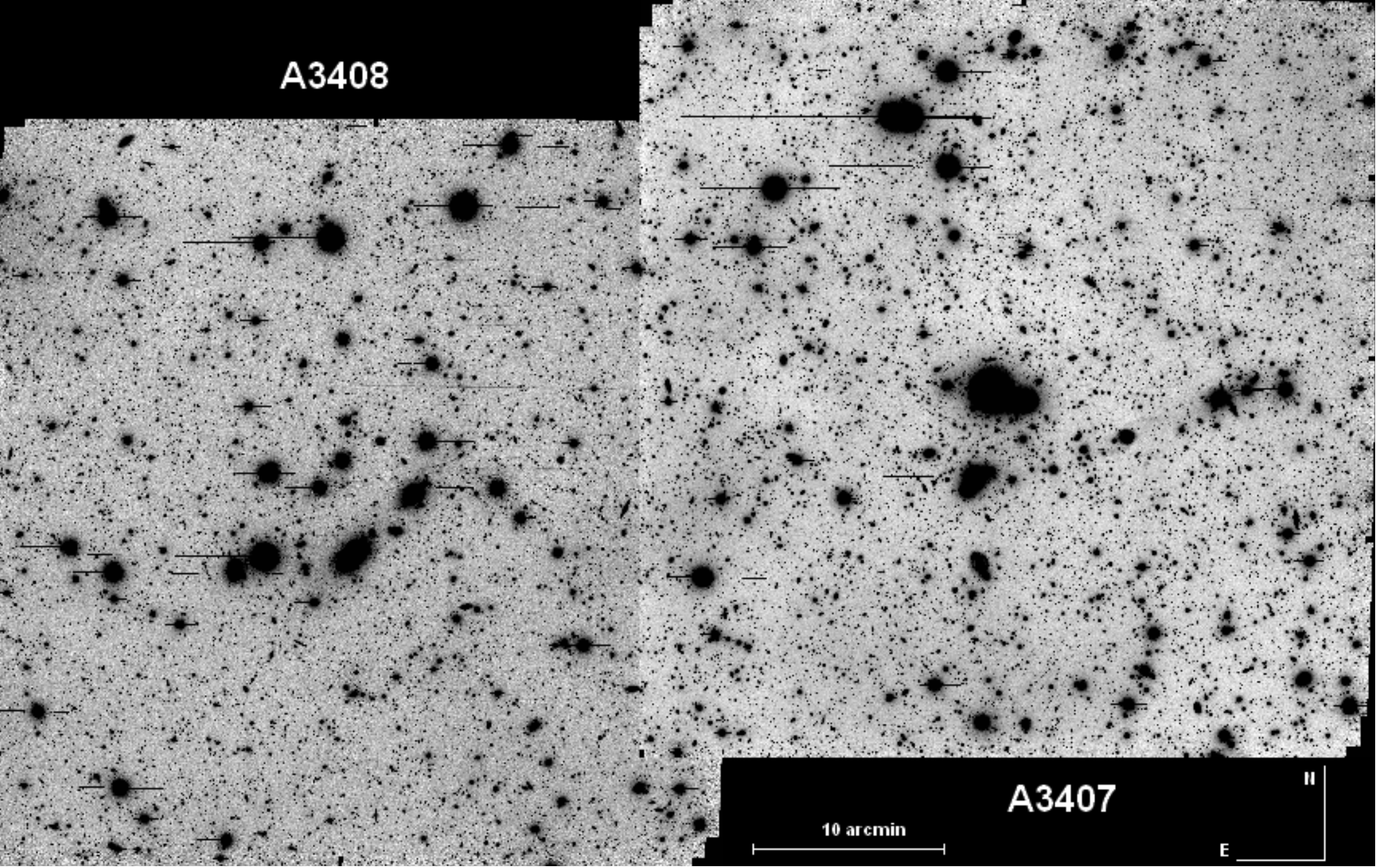}
\caption{Pre-images fields of A3407+A3408 in the R band. The total field is around 73 arcmin.}
\label{fig1}
\end{figure*}

All images and spectroscopic data  of Abell 3407 and Abell 3408 were collected with the 4 meter Victor Blanco telescope at the Cerro Tololo Interamerican Observatory (CTIO), in Chile. The clusters were imaged through the B, V and R Johnson-Cousins filters with the Mosaic II CCD imager during the nights of February 14 and 15, 2007. The Mosaic II array is composed by eight $2048 \times 4096$ SITe CCDs.  With a pixel size of 15 $\micron$ and a scale of 0.27\arcsec/pix, the Mosaic II array cover an area of  $\sim$ 38 arcmin$^2$ on the sky (equivalent to about $1.8 \times 1.8$ Mpc$^{2}$ at the distance of the Abell 3407/Abell 3408 clusters). For Abell 3407, a total of $7 \times 300$ seconds exposures in R-filter and $3 \times 300$ exposures in B and V filters were obtained, given an effective exposure times of 2100 sec for R filter and 900 sec for B and V filters respectively. For Abell 3408, a total of $4 \times 300$ seconds exposure in all three filters were obtained, given an effective exposure time of 1200 seconds for all filters. Abell 3407 was observed under photometric conditions while Abell 3408 was observed under non-photometric conditions (patchy cloudy).  The two clusters were observed with a small overlap  ($\sim 1$\arcmin) to allow to calibrate the observations of Abell 3408. The seeing conditions were poor-to-average in both nights, with a seeing that varied  between 0\farcs9 to 1\farcs6 ({\tt DIMM} monitor). Offsets between  exposures were used to take into account the gaps between the CCDs and for calibration errors. 
In Figure \ref{fig1}, we show the pre-images fields of A3407+A3408 in the R band. The total field is around 73 arcmin.

The optical spectra of objects in Abell 3407 and 3408 were obtained with the Hydra-CTIO multi-object spectrograph \citep{Bar98} on 12 April 2007, during dark time, with a good transparency, and with a seeing that varied  between 0\farcs5 to 0\farcs8 ({\tt DIMM} monitor). The spectra were acquired using the KPGL2 grating over the wavelength range 3450--8242~\AA, centered in 5845~\AA, which provided a spectral resolution of $\sim 6.5$~\AA, and a dispersion of 2.33~\AA~pixel$^{-1}$. To avoid second order contamination above 8000~\AA, the blocking filter GG385 was used. All spectra were imaged with the 400 mm Bench Schmidt camera onto a SITe 2k $\times$ 4k CCD, with a binning of 2 pixels in the spectral direction. Total exposure times of $3 \times 1800$ seconds and $3 \times 1500$ seconds were used for the objects observed in the region of Abell 3407 and Abell 3408, respectively.

\subsection{Data reduction}

The observations were processed with the {\tt MSCRED} package inside IRAF\footnote{IRAF is  distributed by NOAO, which is operated by the Association of Universities for Research in Astronomy Inc., under cooperative agreement with the National Science Foundation.}. The images were bias/overscan-subtracted, trimmed and flat-fielded. The processed images were then registered to a common pixel position and median combined. Calibration on the standard B, V and R magnitude system for Abell 3407 was achieved using observation of stars from \citet{Lan92}. Stars in the overlap region between the two clusters were used to calibrate the photometry in the field of Abell 3408. The accuracy of the calibrations was of the order of 5\% and 7\% for B, V and R filters, respectively

The galaxies used for spectroscopic follow-up were selected using the (B-R) vs R color$-$magnitude diagram. We performed selection of the targets using the Source Extractor software version 2.5 \citep{Ber96}, for both B and R images, taking into account only objects whose galaxy-star separation was 0.3. Furthermore, we selected the targets which are in color range 0 $<$ B-R $<$ 2.5 and brighter than R=20 from color magnitude diagram, (B-R) vs R.

The spectroscopic observations were reduced using the standard procedures in IRAF. All science exposure, comparison lamps (He-Ne-Ar), spectroscopic flats and ''milk-flats'' were bias/overscan subtracted and trimmed using the {\tt CCDRED} package. The ''milki-flats'' were combined and spectral shapes in x- and y- direction were removed using the task {\tt FIT1D}. The resultant image was then filtered by using a median filter and normalized to one. The science exposures and spectroscopic flats were then divided by the processed ''milk flats'' in order to reduce spectral noise in the images.

Cosmic rays removals was performed in the 2D-processed images using the Laplacian Cosmic Ray Identification program \footnote {http://www.astro.yale.edu/dokkum/lacosmic/} \citep{Van01}. The spectra were extracted with {\tt DOHYDRA} task inside the {\tt IRAF HYDRA} package. Dome flats were used to flat field the individual fibers, while twilight flats were used for fiber-to-fiber throughput correction. The spectra were then wavelength calibrated. The residual values in the wavelength solution for 20-30 points using a fourth or fifth order Chebyshev polynomial typically yielded {\em rms} values of $\sim 0.20-0.50$~\AA. Finally, the  average sky spectrum was subtracted from each object spectrum using typically 12 sky fiber spectra per field.

\subsection{Radial velocities}

The radial velocities were determined with the {\tt IRAF RVSAO} package \citep{Kur98}. The task {\tt EMSAO} was used to compute the redshifts for spectra dominated by emission lines. For each identified line, a Gaussian profile is fitting and the radial velocity is computed. Then, the final radial velocity is determined by combining them into a single value. The spectra with observed absorption lines were correlated with 12 high signal-to-noise (S/N) stellar and galaxy templates from \citet{Car06} using the task {\tt XCSAO}. The final heliocentric radial velocities and the $R$ parameter \citep{Ton79}, which gives the quality of the spectra are listed in Tables \ref{Tab_A3407} and \ref{Tab_A3408}.

\begin{figure}
\centering
\includegraphics[width=94mm]{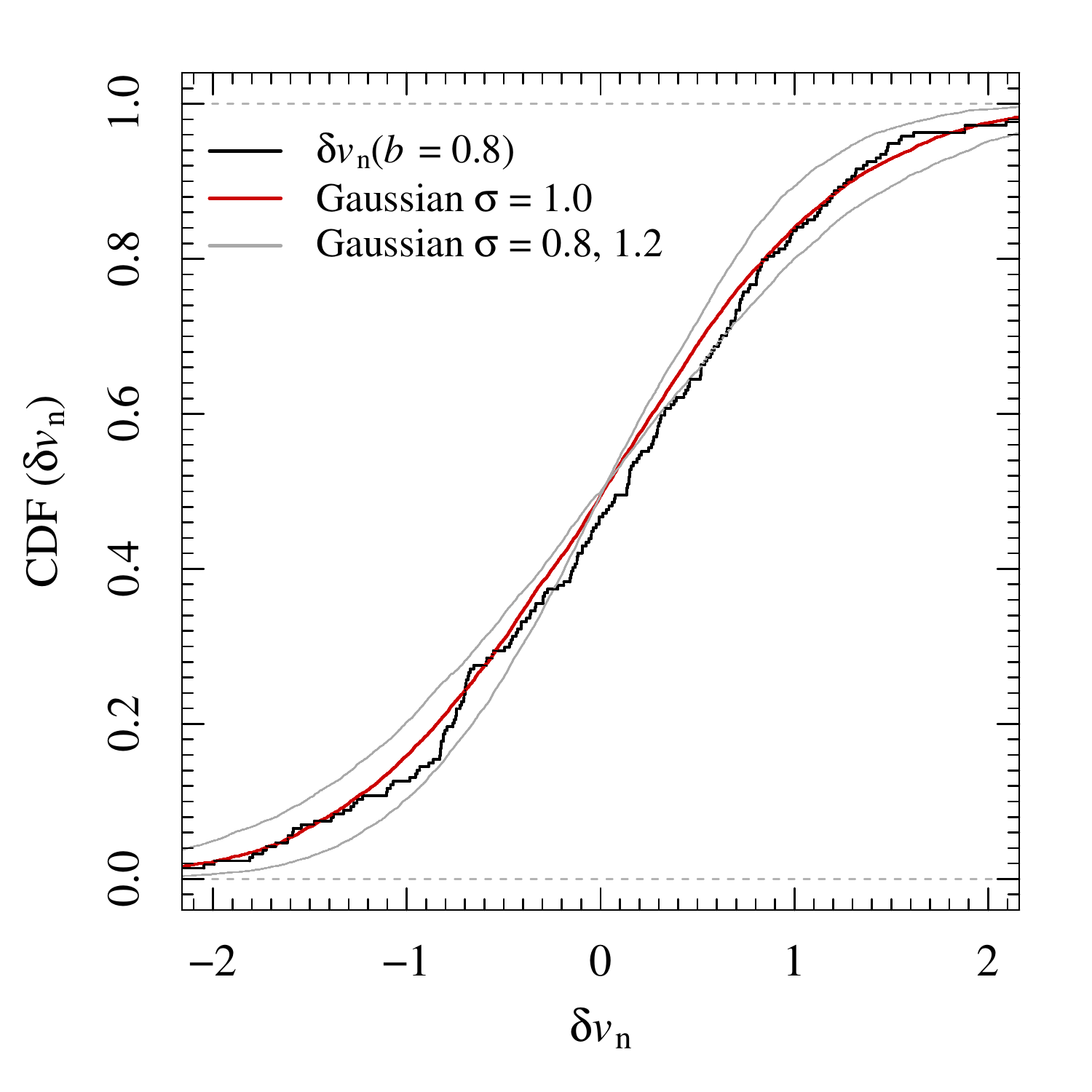}
\caption{Cumulative distribution function (CDF) for the velocity shifts between measurements from different exposures, normalized by the {\tt XCSAO} errors corrected by a factor $b$ ($\delta v_n$, eq. \ref{eq:delta_vn}). For $b = 0.8$, there is a match between the CDF for this quantity and for the Gaussian distribution with $\sigma = 1.0$ (black and red lines, respectively). The gray lines indicate the CDF for the Gaussian distributions with $\sigma = 0.8$ and $\sigma = 1.2$. }.
\label{fig2}
\end{figure}

The following sources of uncertainties were taken into account in the velocity errors: the wavelength calibration errors; the internal error, which accounts for noise in the spectra; and the external error, introduced during the cross-correlation procedure. The first was determined from the wavelength solution, and the second was estimated from the dispersion in velocities obtained with different templates. The last, the external error, corresponds to the error returned by the {\tt XCSAO} task corrected by a factor $b$, which was obtained as follows. Since there are no systematic velocity shifts between different exposures, measurements from different observations of the same object can be used to obtain the calibration factor. The normalized velocity shift is defined as

\begin{equation}
\centering
\delta v_n = \frac{v_1 - v_2}{\sqrt{b^2 (\sigma_1^2 + \sigma_2^2)}}
\label{eq:delta_vn}
\end{equation}

\noindent where $v_1$, $v_2$ are the velocities determined from two different exposures, and $\sigma_1$, $\sigma_2$ are the errors returned by the cross-correlation program. The calibration factor $b$ was obtained assuming that the distribution of the quantity $\delta v_n$ must be Gaussian with dispersion 1.0 and folded about zero. Using the Kolmogorov-Smirnov test, we compared the distribution of the normalized velocity shifts, $\delta v_n$, and the folded Gaussian.  A satisfactory match between these two distributions is achieved if the errors are multiplied by $b = 0.8$ (Figure \ref{fig2}). This correction factor was applied to the {\tt XCSAO} errors. The final error ${\rm \delta v_{final}}$ was obtained by adding in quadrature each error term. On average, we find $\langle{\rm \delta v_{final}\rangle\approx 53}$ ${
\rm km~s^{-1}}$.

\section{Velocity distribution}

Combining our spectroscopic observations with the data available at the 6dFGS\footnote{6dFGS Database http://www-wfau.roe.ac.uk/6dFGS/} database \citep{JRS}, we gathered radial velocities for 156 galaxies in the field of $\sim 3^\circ \times 3^\circ$ centered at the mid-distance between A3407 and A3408 (see Figure \ref{fig3}). Of this total, there are 21 galaxies, identified as repeated objects, which were used to check  consistency between the two redshift surveys. Computing the absolute difference between the two independent radial velocity measurements, we find $\langle{\rm |\Delta V|\rangle\approx 45}$ ${\rm km~s^{-1}}$ on average (see Figure \ref{fig3}). This quantity is a little smaller than $\langle{\rm \delta v_{final}\rangle}$, indicating that our combined sample is internally consistent.

\subsection{Membership, location and scale}

All properties of galaxy clusters can be significantly affected by projection effects. Over the years, many methods have been developed to remove interlopers from galaxy clusters \citep[e.g.][]{Yah77,DK,FGG}. This introduces the problem of picking the best method in each situation. These methods show little differences in final results, mostly coming from borderline galaxies which do not significantly contribute to bias the cluster properties \citep[see][]{WO08}.

\begin{figure}
\centering
\includegraphics[width=94mm]{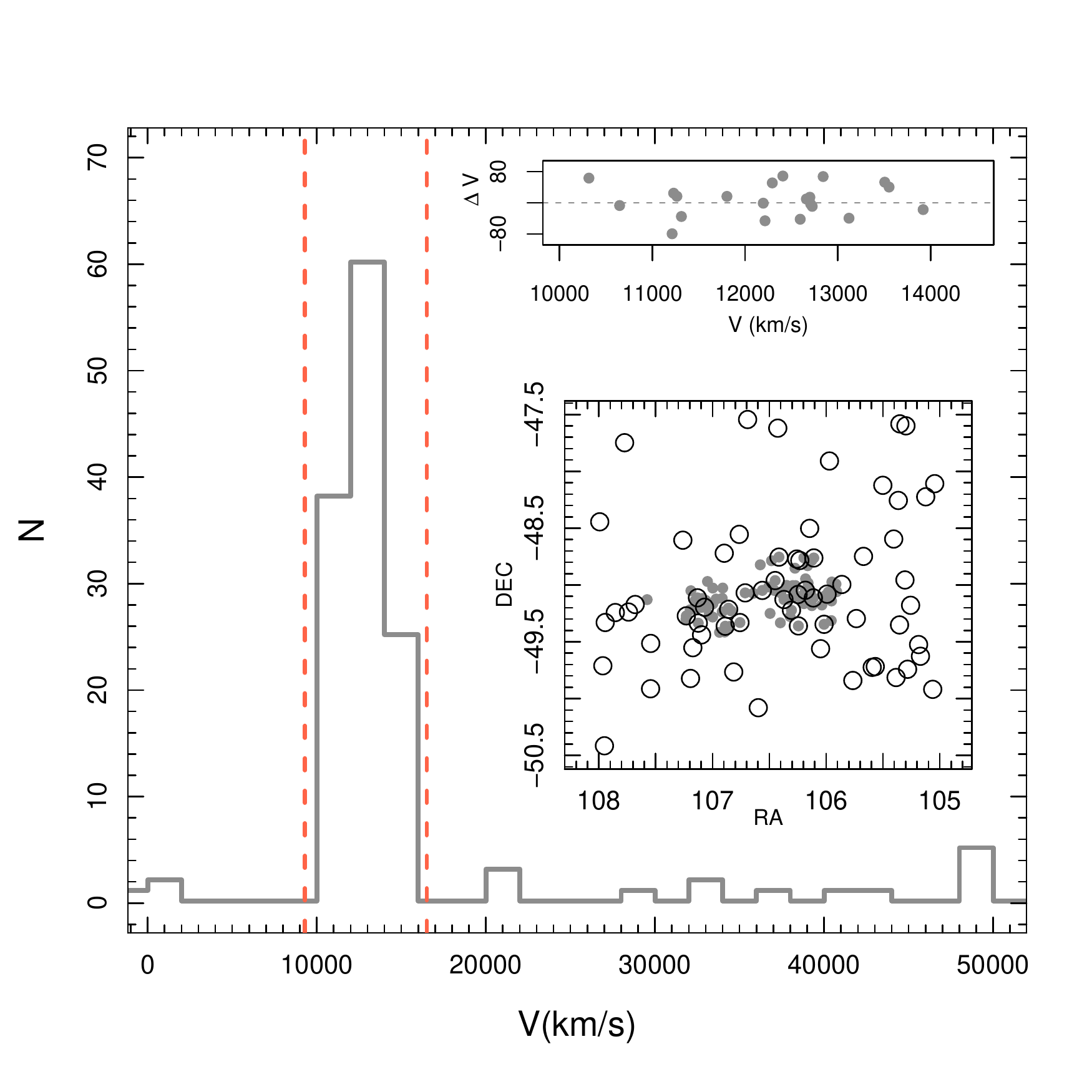}
\caption{ Velocity distribution before discarding outliers. Red dashed lines show the initial cut-off of $\pm$ 3000 ${\rm km~s^{-1}}$ around the median redshift, ${\rm z_{cl}\approx 0.042}$.
The secondary panel shows the spatial distribution of our survey (gray filled circles) plus the 6dF data (open circles). The small upper panel shows the velocity comparison of the 21 objects common to our sample and 6dF.}
\label{fig3}
\end{figure}

\begin{figure}
\centering
\includegraphics[width=94mm]{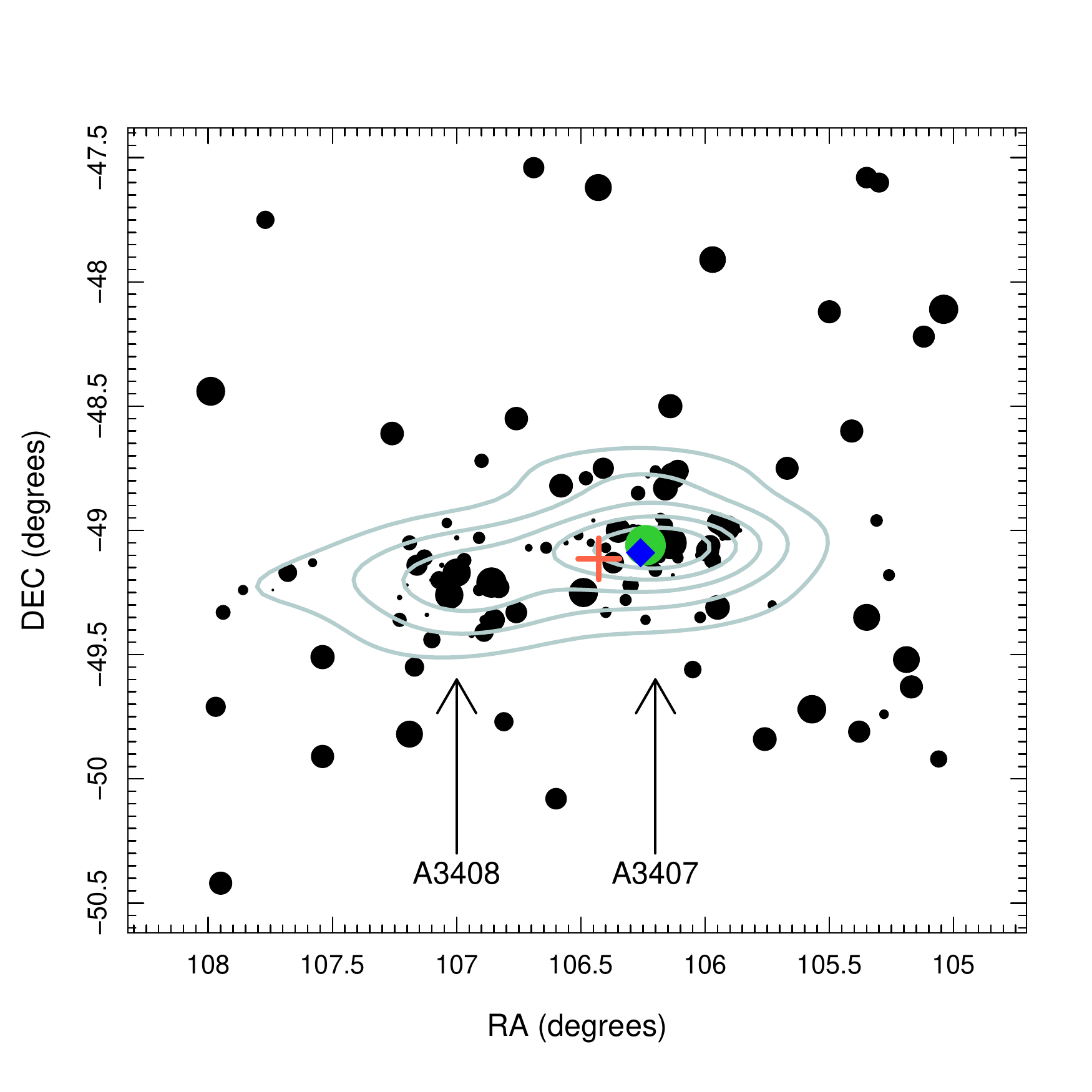}
\caption{Galaxy distribution around A3407 and A3408. The size of each point is proportional to the galaxy luminosity. The central part of the field is outlined by the density contours. The red cross indicates the luminosity-weighted centroid of the distribution, the green filled circle indicates the BCG, and
the blue diamond is the peak of the X-ray emission.}
\label{fig4}
\end{figure}

In this work, we apply an initial cut-off of $\pm$ 3000 ${\rm km~s^{-1}}$ around the cluster redshift, ${\rm z_{cl}\approx 0.042}$, selecting 125 galaxies in the approximate range $9800 \lesssim {\rm V} \lesssim 15300$ ${\rm km~s^{-1}}$ (see Figure \ref{fig3}). Additionally, we use the shifting-gapper method \citep{FGG} to reject remaining interlopers. Here, we follow the procedure outlined by \citet{OW09,ONC}. The initial step is to determine the center of the galaxy spatial distribution. Although we have two nominal clusters in the field, we first consider that A3407 and A3408 may form a single system, at least in the velocity space, and hence we need to provide one single center to this field. Using Rosat-All-Sky-Survey maps, we define this center as the peak in the X-ray emission from A3407, the richest cluster of the pair. The brightest cluster galaxy (BCG) of A3407, ESO 207-19, an object of absolute magnitude $M_R=-22.64$, is $\sim 51^{\prime\prime}$ ($\sim 30$ kpc) away from this peak, and $\sim 7^\prime$ ($\sim 240$ kpc) away from the luminosity-weighted centroid of the distribution (see Figure \ref{fig4}).

Next, galaxies are sorted into bins as a function of radial distance from the center of the cluster. The bin size is 0.4 Mpc or larger to force the selection of at least 15 galaxies \citep{FGG,L09}. Within each radial distance bin, galaxies are sorted by their peculiar velocity with respect to the velocity of the cluster. We define this peculiar velocity as:

\begin{equation}
 v_{pec}^{i}= c{(z_i - \bar{z})\over (1+ \bar{z})}
\end{equation}

\noindent where $v_{pec}^{i}$ is the peculiar velocity of galaxy $i$, $z_i$ is the redshift of galaxy $i$, and $\bar{z}$ is the average redshift of the cluster. In each bin, the ``f-pseudosigma`` \citep{Bee90} is determined and used as the velocity gap to reject outliers. The value of f-pseudosigma ($S_f$) corresponds to the normalized difference between the upper ($F_u$) and lower ($F_l$) fourths of a data set. It can be calculated as follows:

\begin{equation}
S_f = {(F_u - F_l)\over 1.349}
\end{equation}

\noindent  The constant 1.349 is the typical difference ($F_u - F_l$) for standard normal distributions \citep{Bee90}. This process is repeated for each bin until either the number of sources stabilizes, the value of f-pseudosigma drops below 250 ${\rm km~s^{-1}}$, or the value of f-pseudosigma begins to increase \citep[e.g.][]{WB}.

\begin{figure}
\centering
\includegraphics[width=94mm]{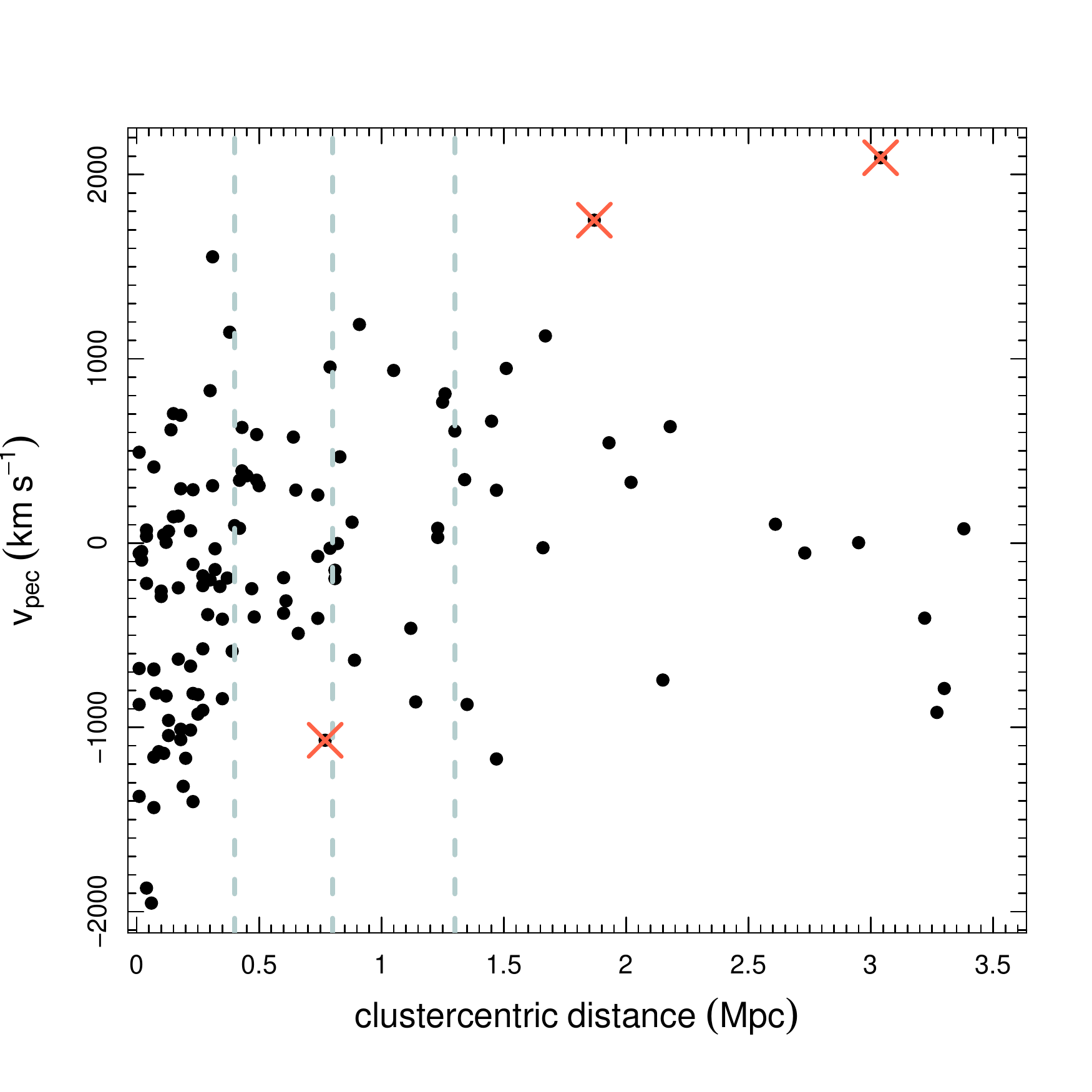}
\caption{Galaxy clustercentric distances against the peculiar velocities with respect to the cluster velocity.
The size of each point is proportional to the galaxy luminosity.
The objects marked in red are those excluded by the outliers removal process. The vertical dashed lines indicate the radial bins within which galaxies were sorted by their peculiar velocity.}
\label{fig5}
\end{figure}

After completion of the removal process, a total of 122 galaxies remained in the sample within a clustercentric radius of $\sim$ 3.5 Mpc (see Figure \ref{fig5}).\footnote{The result of the removal process is the same as choosing the centroid or the BCG location as the center of the system.} For this data set we determine the location (velocity mean) and scale (velocity dispersion), using the biweight estimators ${\rm C_{BI}}$ and ${\rm S_{BI}}$ \citep[see the definitions in][]{Bee90}. These estimators have a wide use in science since they are less sensitive to outliers and the shape of the underlying distribution \citep[e.g.][]{CD}. By applying the ROSTAT software \citep{Bee90}, we find ${\rm C_{BI}}=12415_{-50}^{+110}$ ${\rm km~s^{-1}}$, and ${\rm S_{BI}}=691_{-35}^{+74}$ ${\rm km~s^{-1}}$. The errors correspond to the 68\% confidence interval, calculated after bootstrap resamplings of 10,000 subsamples of the velocity data. 

\begin{figure}
\centering
\includegraphics[width=94mm]{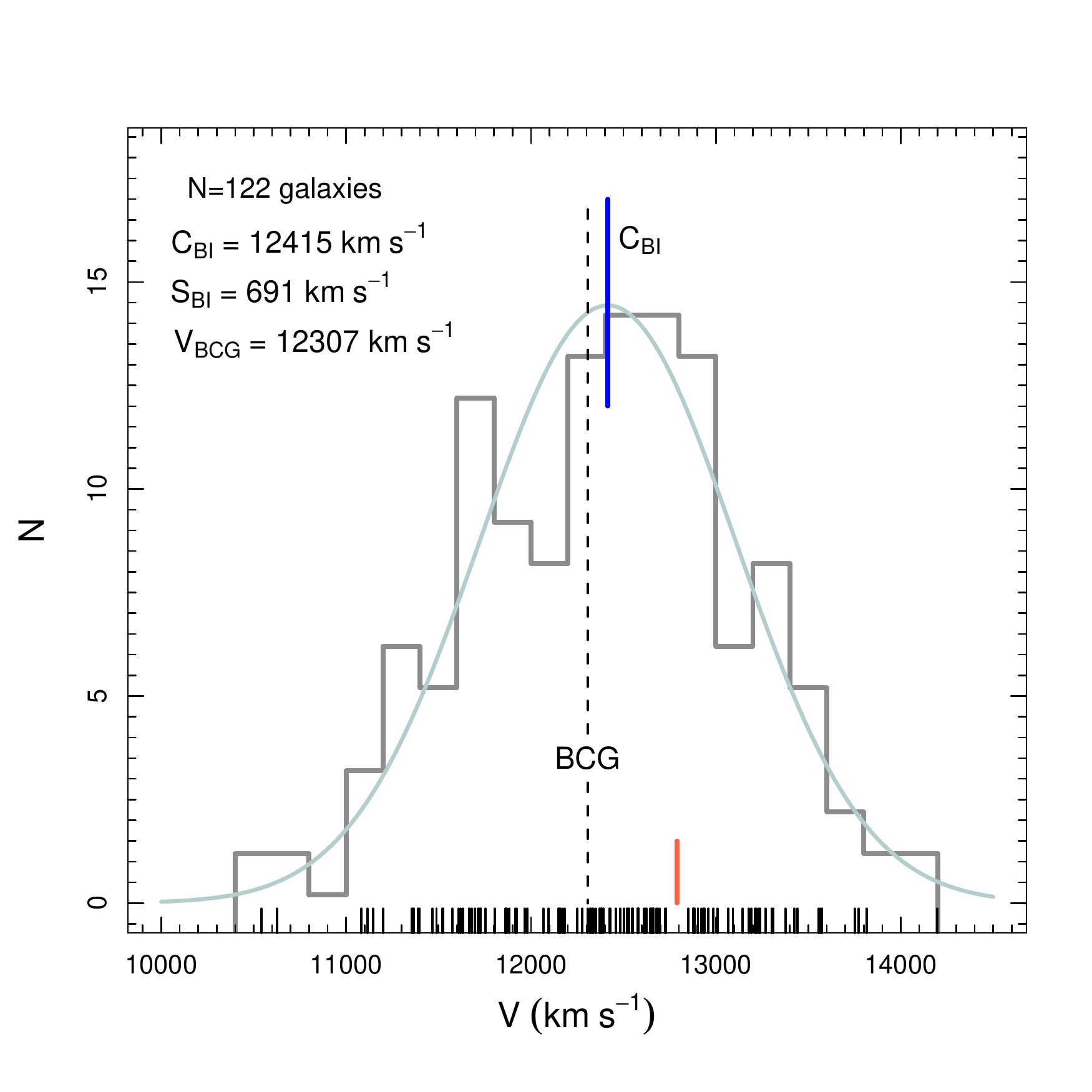}
\caption{Velocity distribution histogram with bins of width 200 ${\rm km\;s^{-1}}$. The solid line is the Gaussian with mean 12415 ${\rm km\;s^{-1}}$ and standard deviation of 691 ${\rm km\;s^{-1}}$.
The vertical blue line indicates the mean and the dashed line indicates the BCG. At the bottom, a rug plot is shown, with the red line indicating the position of the significant weighted gap.}
\label{fig6}
\end{figure}

The velocity distribution of the clusters members is shown in Figure \ref{fig6}. Also in this figure we depict the position of the BCG galaxy , ${\rm V_{BCG}=12307\pm 24}$ ${\rm km~s^{-1}}$. Following \citet{TCG} we test if this galaxy can be central in the A3407 + A3408 velocity field using

\begin{equation}
{\rm S_V = {|C_{BI} - V_{BCG}|\over ~~ [\delta v_{cl}^2 + \delta v_{BCG}^2]^{1/2}}},
\end{equation}

\noindent where ${\rm \delta v_{cl}}$ and ${\rm \delta v_{BCG}}$ are the errors in the biweight and the BCG velocities. Values of ${\rm S_V \gtrsim 2}$ mean the BCG peculiar motion is probably significant. For our sample, ${\rm S_V \simeq 1.29}$, indicating that we can not reject the BCG location as being on the dynamical center of the field (see Figure \ref{fig6}). Indeed, the BCG is offset relative to the X-ray peak by only $\sim 30$ kpc, and offset scales up to $\sim 50$ kpc can be explained by small amplitude oscillations of the central galaxy around the bottom of the cluster potential well \citep[see][]{LC}.

\subsection{Normality}

The  velocity distribution of galaxies in clusters can provide information about the dynamical state of these systems. The normality of the radial velocity distribution is usually related to the dynamical equilibrium of a galaxy cluster. Both theoretical and phenomenological developments suggest that the virialized equilibrium state of a spherical gra\-vi\-ta\-ti\-o\-nal system is approximately described by a Maxwell-Botzmann distribution function \citep[][]{OG57,LY67,UIS,HW,BW12,BLS}. In phase-space, this translates to a Gaussian function (or Normal distribution). N-body numerical experiments of the relaxation of single isolated gravitational systems \citep{MH} or that of cosmological halos \citep{HEH,HMZ} also support these conclusions. 

This suggests that discriminating groups according to their velocity distributions may be a promising way to assess the dynamics of galaxy systems. Unfortunately, this is not a simple task. \citet{Bee90} stress the difficulty in determining when a given velocity distribution differs significantly from normality, pointing out that the classification of a cluster as Gaussian or non-Gaussian may be dependent on the statistical test used in the analysis. This suggests the need of using several complementary statistical tests to achieve a reliable diagnostic \citep{Bee90,BB,Hou09,RDT,RS}.

\begin{table}
\begin{center}
{\small
\caption {Statistical tests for normality with the respective p-values and diagnostics at the 95\% confidence level.} 
\label{tab1}
\begin{tabular}{|l c c|}
\hline
TEST & P-VALUE & DIAGNOSTICS\\
\hline\hline
${\rm W^2}$ & 0.276 & NORMAL\\
${\rm U^2}$ & 0.258 & NORMAL\\
${\rm A^2}$ & 0.414 & NORMAL\\
KS          & 0.250 & NORMAL\\
\hline
B2          & 0.330 & NORMAL\\
TI          & 0.113 & NORMAL\\
a           & 0.218 & NORMAL\\
W           & 0.678 & NORMAL\\
u           & 0.800 & NORMAL\\
\hline
B1          & 0.278 & NORMAL\\
AI          & 0.193 & NORMAL\\
\hline
\end{tabular}
}
\end{center}
\label{T1}
\end{table}

We can distinguish three categories of normality tests among those included in the ROSTAT package. The omnibus tests, which try to quantify the overall deviation of the velocity distribution from a Gaussian, such as the Cramer von-Mises ${\rm W^2}$ test, the Watson ${\rm U^2}$ test, the Anderson-Darling ${\rm A^2}$ test, and the Kolmogorof-Smirnov (KS) test \citep[see][for references]{BGF}. The shape tests, which are devised to measure the shape of the outskirts of the distribution, such as the kurtosis test (the B2 test) and its robust counterpart, the Tail Index (TI) test \citep[see][for a discussion]{BB}, or to test its tail population, such as the a and the W tests, which are most sensitive to the tail of the underlying populations, and the u test, which is sensitive to contamination by extreme values \citep[see][for a discussion on these tests]{Yah77}. Finally, there are tests which measure the asymmetry of the distribution: the skewness test (B1 test) and its robust version, the Asymmetry Index (AI) test \citep{BB}. For each of these tests, ROSTAT computes its statistics as well as their associated p-values \citep{Bee90}. In Table \ref{tab1} we present the results of all these tests, which  unanimously failed to reject he normality of the velocity distribution of the A3407 + A3408 field.

\subsection{Unimodality}

Although the tests used in the previous section consistently indicate normality, we explore the possibility of gaps in the velocity distribution. The ROSTAT package provides two statistical tests helping to identify kinematical features in the velocity distribution. These are the gap analysis \citep{WS} and the dip  unimodality test \citep{HH}. The dip test measures the maximum difference, over all sample points, between the empirical distribution function, and the unimodal distribution function. Since the dip statistic is asymptotically larger for the uniform than for any distribution in a wide class of unimodal distributions, it appears to be a reasonable measure of the extent of deviation from unimodality \citep{HH}. The gap analysis estimates the probability that a gap of a given size and location, between the ordered velocities, may be produced by random sampling from a Gaussian population. First, the velocities are sorted in increasing order and the $i$th velocity gap is given by $g_i=v_{i+1} - v_i$. The weight for the $i$th gap is $w_i=i(N - i)$ and the weighted gap is defined as $\sqrt{w_ig_i}$. The weighted gaps are normalized through dividing by the mid-mean ${\rm (MM)}$ of the ordered weighted gap distribution given by

\begin{equation}
{\rm MM}= {2\over N}\sum_{i=N/4}^{3N/4} \sqrt{w_ig_i}.
\end{equation}

\noindent We look for normalized gaps larger than 2.25, since in random draws of a Gaussian distribution they arise at most in $\sim$ 3\% of the cases \citep[see][]{WS,BGF}. We detect one significant gap at ${\rm V \simeq 12790~km~s^{-1}}$ (see Figure \ref{fig6}). This could be an indication that the distribution is bimodal. However, normality was not rejected after a battery of tests, and the dip test also failed to reject the unimodality with p-value=0.7542. To explore a little more this result, we test if our sample can be model as a normal mixture  using  the Mclust code \citep{FR}. Mclust is a contributed R package,  an  open-source  free  statistical  environment  developed  under  the  GNU  GPL  \citep[][http://www.r-project.org]{IG}. The method is based on a search of an optimal model for the clustering of the data among models with varying shape, orientation and volume. It finds the optimal number of components and the corresponding classification (the membership of each component). We run Mclust on  10,000 bootstrap resamplings of the velocity data and on 10,000 realizations of a normal distribution with $\mu={\rm C_{BI}}$ and $\sigma={\rm S_{BI}}$. We find bimodality in 13\% of times for the resamplings of the observed distribution, and 11\% of times for the normal realizations. The closeness of these values suggests a situation where bimodality may be undetectable to Mclust and other statistical tests. We should recall that not all mixture of two unimodal distributions with differing means is necessarily bimodal. For instance, the \citet{HV} bimodality indicator, given by

\begin{equation}
d={|\mu_1 - \mu_2|\over 2\sqrt{\sigma_1\sigma_2}},
\end{equation}

\noindent indicates that a mixture should be considered bimodal only if $d> 1$. For the A3407$+$A3408 velocity field, the mixture detected with Mclust has the following parameters: $\mu_1\simeq 11979~{\rm km~s^{-1}}$, $\sigma_1\simeq 596~{\rm km~s^{-1}}$, $\mu_2\simeq 12826~{\rm km~s^{-1}}$, and $\sigma_1\simeq 527~{\rm km~s^{-1}}$, which leads to $d=0.75$, and thus we can not reject unimodality.

\begin{figure}
\centering
\includegraphics[width=94mm]{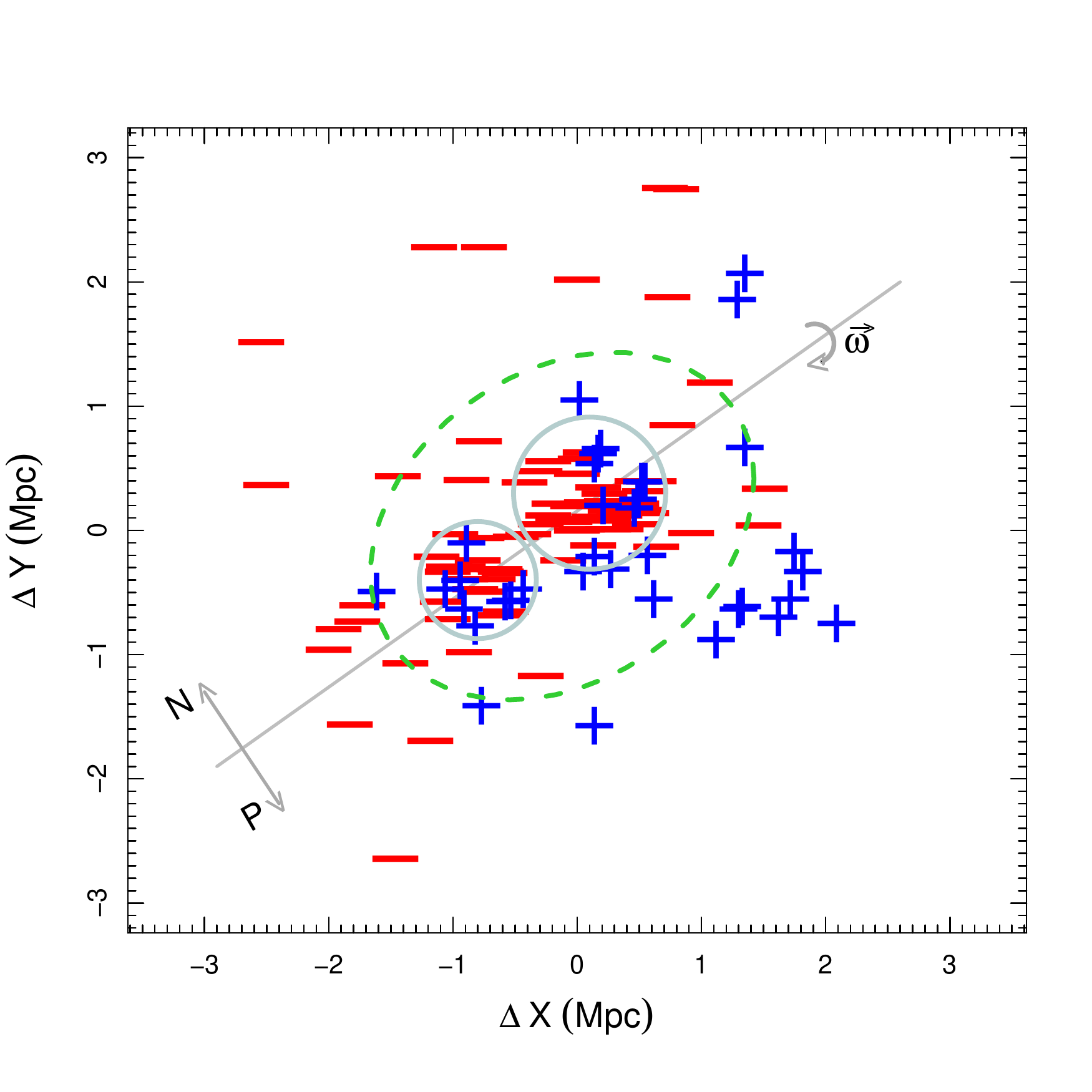}
\caption{Distributions of galaxies projected on the sky plane. The symbols '$+$' and '$-$' indicate positive or negative velocity with respect to the gap position. The vector $\vec{\omega}$ indicates
the possible rotation around the principal axis. Directions used for distance summations are given by 'N' (negative) and 'P' (positive). The dashed green ellipse indicates the region containing 90\% of data. 
The circles indicate the approximate regions of A3407 and A3408.}
\label{fig7}
\end{figure}

\subsection{Velocity gradient}

Up to this stage, we found that the velocity distribution of the A3407+A3408 field is consistent with both unimodality and normality. Also, the BCG galaxy does not have a significant peculiar motion with respect to the center of the velocity distribution. Now, we want to consider a further question: can the  gap identified in the velocity distribution be indicating a velocity gradient across the spatial galaxy distribution? To explore this possibility, we estimate the principal axis of the galaxy projected distribution using the moments of inertia method \citep{CM}. From the eigenvectors of the spatial configuration, we find the direction of the principal axis of the system, as shown in Figure \ref{fig7}. 
Next, we divide the sample in two groups, above or below the gap position, and compute the perpendicular distance of each object to the principal axis. We adopt arbitrary signs for the distances on each side of the axis (see Figure \ref{fig7}). Let us call S1 and S2 the total sum of distances for objects in group 1 (${\rm V < 12790~km~s^{-1}}$ \& 86 galaxies) and in group 2 (${\rm V > 12790~km~s^{-1}}$ \& 36 galaxies). We find ${\rm S1=-11.92}$ Mpc and ${\rm S2=14.60}$ Mpc, respectively. The significance of this result is determined by running 10,000 realizations of two spatial Poisson processes with the same number of points as groups 1 and 2, and within the same limits of the real galaxy distribution: $[-3,3]~\times ~[-3,3]~{\rm Mpc}$. Distances of all points to the principal axis are computed and summed just as we did before. At the end of the runs we have an output distribution reflecting the possible range of sums. To achieve a result indicating a significant velocity gradient, at least 
one of the sums observed (S1 or S2) must be outside the robust 95\% confidence level interval, which are ${\rm I1}:[-12.53,19.86]$ Mpc and ${\rm I2}:[-10.32,13.05]$ Mpc (for Poisson processes with 86 and 36 points, respectively). Hence, while S1 is consistent with I1, S2 is more positive than I2, suggesting the possibility of an asymmetric velocity distribution across the spatial distribution, i.e., high velocity galaxies may be segregated spatially with respect the principal axis. This velocity gradient of $\sim 847\pm 114~{\rm km~s^{-1}}$ could indicate some rotation $\vec{\omega}$
around this axis (see Figure \ref{fig7}). It is consistent with typical gradients (240 $-$ 1230 ${\rm km~s^{-1}}$) found in clusters studied by \citet{DK}. 

A galaxy system can acquire angular momentum from cosmological {\it ab initio} conditions from their formation times \citep{LI98, LI14}. Another possibility is through an off-axis merging between two clusters \citep{RI98,TA00,PED}. This does not seem to be the case of A3407 \& A3408, since they both are well aligned with the cluster principal axis (see Figure \ref{fig7}).
It should also be noted in this figure that the two velocity components (the positive and negative values with respect to the gap) are widespread in the field, permeating both A3407 and A3408. That is, the velocity gradient is not reflecting only the central structures in the field. This could mean that the galaxy distribution in and around A3407 \& A3408 may have acquired this pattern in the same cosmic event, whose nature is not clear at the moment.

Finally,  it is worth noting that a rotating system does not mean a nonequilibrium system. In fact, \citet{HL} studied two probable rotating clusters (Abell 954 and Abell 1399) and verified that they may be in dynamical equilibrium and have undergone no recent merging. Similarly, \citet{Oeg92} found that the highly probable rotating cluster Abell 2107 has galaxy velocities consistent with a Gaussian distribution.

\section{Subclustering}

Although the A3407$+$A3408 field can be described by a single Gaussian velocity distribution, we can not say they constitute a single dynamical unit without taking into account their spatial coordinates. To examine this more general situation, we consider once more that A3407 and A3408 may form a single cluster -- the hypothesis to be tested. Then, we apply statistical tests to check if A3407 and A3408 emerge as independent entities.

A cluster is said to contain substructures (or subclusters) when its surface density is characterized by multiple, statistically significant peaks, in combination with the distribution of galaxy velocities \citep[e.g.][]{RA07}. A variety of statistical tests are available to assess the presence of substructures in galaxy clusters \citep[see][]{Pin96,BMB}. We chose to apply four of them: The $\beta$ test \citep{Wes88}, the $\Delta$ test \citep{Dre88}, and the Lee 2D and 3D statistics \citep{Lee79,Fic87}, conducted here following the work of \citet{Pin96}. They can be briefly described as:

\begin{itemize}
\item The $\beta$ statistics is a two-dimensionional estimator of deviations from the mirror symmetry about the cluster center. 
\item The $\Delta$ statistics evaluates the kinematics of galaxy groups identified in sky projected clusters. 
\item The Lee 2D statistics  is a measure of the clumpiness in the locations of galaxies after they have been projected onto a line. 
\item The Lee 3D statistics extends the procedure to include a third "dimension" given by velocity data. 
\end{itemize} 

Results presented in Table \ref{tab2} indicate the presence of subclusters in the field. To separate them, we take the full available phase-space information making use of the KMM algorithm \citep{ABZ}, applied to the distribution of cluster members in the 3D-space of positions and velocities. We search for the solution that separates the members into two subclusters. The KMM algorithm uses the maximum-likelihood ratio test to estimate how likely the two-system solution is to be a significant improvement over the single-system solution \citep[e.g.][]{Ba2}.

\begin{table}
\begin{center}
{\small
\caption {Statistical tests for substructures with the respective p-values and diagnostics at the 95\% confidence level.} 
\label{tab2}
\begin{tabular}{|l c c|}
\hline
TEST & P-VALUE & DIAGNOSTICS\\
\hline\hline
$\beta$ & 0.001 & \scriptsize{SUBSTRUCTURES}\\
$\Delta$ & 0.028 & \scriptsize{SUBSTRUCTURES}\\
Lee 2D & 0.002 & \scriptsize{SUBSTRUCTURES}\\
Lee 3D& 0.016 & \scriptsize{SUBSTRUCTURES}\\
\hline
\end{tabular}
}
\end{center}
\end{table}

\begin{figure}
\centering
\includegraphics[width=94mm]{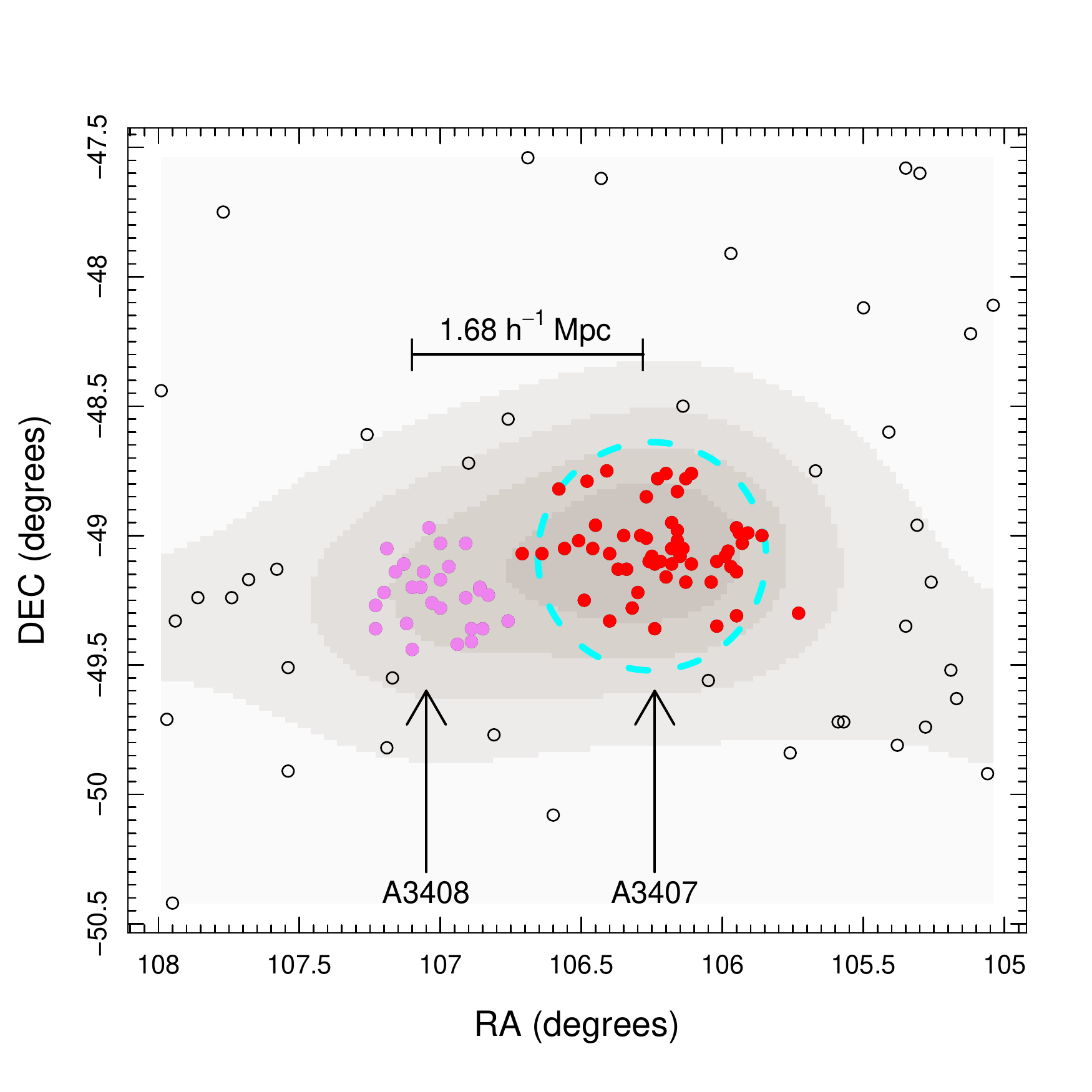}
\caption{Two subclusters, corresponding to A3407 and A3408, identified by the KMM algorithm. Densities are depicted by gray shades. A3407 and A3408 are indicated in red and purple colors. The cyan circle encompasses the region whose density is 2$\sigma$ higher than the average.}
\label{fig8}
\end{figure}

\begin{figure}
\centering
\includegraphics[width=74mm]{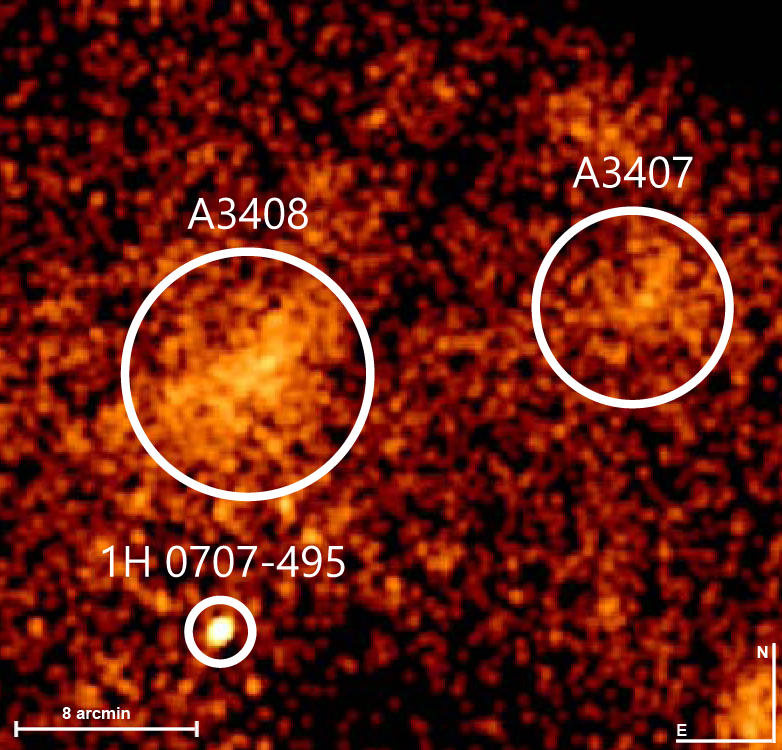}
\caption{ROSAT PSPC smoothed (with a 3'' width gaussian) image from a pointed 9.5 ksec observation RP180306N00. Also shown the AGN 1H 0707-495, the main target of that observation. The circles surrounding A3408 and A3407 have 10' and 8' radii and ilustrate the X-ray emission of the A3407-A3408 pair.}
\label{fig9}
\end{figure}

We find that the solution with two subclusters is significantly better than the single unit solution, at the 99\% confidence level. In Figure \ref{fig8} we show the two subclusters identified. Not surprisingly
they correspond to A3407 and A3408, individually identified for the first time in the present work. In this figure we also indicate the circle containing the region whose density is 2$\sigma$ higher than the average of the full sample (let us call it $R_{2\sigma}$, with $R _{2\sigma}=0.88{\rm h^{-1}}$ Mpc). This circle completely encompasses the main system (A3407), and leaves the secondary system (A3408) completely outside $R_{2\sigma}$. KMM assigns 27 galaxies to A3408, each at the 99\% c.l. From these galaxies, we compute the mean velocity ${\rm V_{\mbox{\scriptsize{{A3408}}}} = 12458\pm 98}$ ${\rm km\;s^{-1}}$ and velocity dispersion ${\rm \sigma_{\mbox{\scriptsize{{A3408}}}} = 573_{-37}^{+48}}$ ${\rm km\;s^{-1}}$. At the same time, KMM assigns 54 galaxies to A3407, each at the 99\% c.l. The resulting mean velocity is ${\rm V_{\mbox{\scriptsize{{A3407}}}} = 12328\pm 116}$ ${\rm km\;s^{-1}}$ with velocity dispersion ${\rm \sigma_{\mbox{\scriptsize{{A3407}}}} = 718_{-42}^{+61}}$ ${\rm km\;s^{-1}}$. The projected distance between the X-ray peaks of each cluster is $1.68~{\rm h^{-1}}$ Mpc. This distance extends beyond $R_{2\sigma}$, suggesting a significant physical separation between the clumps. On the other hand, the relative velocity between the two systems is not significant in comparison with the errors, $\Delta V = 130 \pm 151$ ${\rm km\;s^{-1}}$. This small velocity separation is consistent with our previous findings on normality and unimodality, suggesting the existence of a single cluster in the velocity space. However, the double peak in the galaxy spatial distribution is also present in the X-ray emission, as we can see in the Rosat-All-Sky-Survey image -- Figure \ref{fig9}. In this figure we also see that there is no common X-ray halo in the field, which weakens the idea of a single dynamical unit for A3407+A3408. From now on, we assume that A3407 and A3408 are individual systems.

\section{Dynamical analysis}

\subsection{Virial mass}

Before obtaining the virial mass of each individual cluster, we need to check whether their velocity distributions are Gaussian. If that is the case, we can assume virialization. Applying the same statistical tools used in Section 3.2, we verified that none of them rejects normality of the velocity distributions of A3407 and A3408 at the 95\% c.l. (with p-values $\geq$ 0.116 and 0.347, respectively). Another virialization indicator is the crossing-time of a galaxy system. According to \citet{NW}, a system with crossing-time $>$ 0.09 ${\rm H_0^{-1}}$ (hereafter $t_c^{vir}$) probably has not yet had time to virialize. The crossing-time is calculated as 

\begin{equation}
t_c = {3 r_H\over 5^{3/2}\sigma_v}
\end{equation}

\noindent \citep{HG}, where the harmonic radius, $r_H$ is independent of the velocity dispersion and is given below:

\begin{equation}
r_H = \pi D \sin{\left[{n(n-1)\over 4\sum_i\sum_{j>i} \theta_{ij}^{-1}}\right]},
\end{equation}

\noindent where $D$ is the distance to the group, $n$ is the number of members of each group, and $\theta_{ij}^{-1}$ is the angular separation of group members. Setting up the cosmological parameters as $\Omega_M=0.3$ and $\Omega_\Lambda=0.7$, we find the distances to each system: 122.14 $h^{-1}$ Mpc (A3407) and 123.40 $h^{-1}$ Mpc (A3408), that leads us to $t_c=0.041~{\rm H_0^{-1}}$ (A3407) and $t_c=0.037~{\rm H_0^{-1}}$ (A3408), which is a further indication that these systems are virialized. In the same way, the crossing-time for the whole A3407+A3408 cluster is $t_c=0.053~{\rm H_0^{-1}}$, also a lower value than $t_c^{vir}$, indicating that even this larger system has had time to virialize.

Assuming from now on that the virial theorem applies, the system is self-gravitating, and the bodies in the system have equal masses, the virial mass estimator is usually written as:

\begin{equation}
M_V={3\pi N\over 2 G} {\sum(v_i - V)^2\over \sum_{i<j} 1/R_{ij}},
\end{equation}

\noindent where $N$ is the number of cluster members, $v_i$ is the velocity of the $i$-th galaxy, $V$ is the mean of all members, and $R_{ij}$ is the projected separation between the two galaxies $i$ and $j$ \citep{HTB}. We summarize all the structural and dynamical properties in Table \ref{tab3}. Errors on $r_H$ and $\sigma$ were calculated using the bootstrap method for 10,000 resamplings and then used standard error propagation analysis to calculate the rms error on the crossing-time and the virial mass.

\begin{table}
\begin{center}
{\small
\caption{Structural and dynamical properties of clusters  A3407 and A3408.}
\label{tab3} 
\begin{tabular}{|c c  c  c  c |}
\hline
 Cluster & $\sigma$ & $r_H$ & log $M_{V}$ & $t_c$ \\
  ~~~~   & ( ${\rm km\;s^{-1}}$) & ($h^{-1}$ Mpc) & ($h^{-1}$ ${\rm M_{\odot}}$) & (${\rm H_0^{-1}}$) \\
\hline
A3407 & 718$^{+93}_{-48}$ & 1.10$^{+0.12}_{-0.13}$ & 14.59$^{+0.46}_{-0.42}$ & 0.041$^{+0.011}_{-0.009}$ \\
\\
A3408 & 573$^{+82}_{-59}$ & 0.80 $^{+0.21}_{-0.13}$& 14.26$^{+0.55}_{-0.46}$ & 0.037$^{+0.010}_{-0.009}$\\
\\
A3407+08 & 691$^{+74}_{-35}$ &  1.36 $^{+0.27}_{-0.18}$& 14.66$^{+0.64}_{-0.57}$ & 0.053$^{+0.012}_{-0.012}$     \\
\hline
\end{tabular}
}
\end{center}
\end{table}

\subsection{Two-body model}

The stage is being set for a possible merger of A3407 and A3408. By knowing their virial masses and the spatial separation, we can use the Newtonian binding criterion that a two-body system is bound if the potential energy of the bound system is equal to or greater than the kinetic energy. To assess the likelihood that A3407 and A3408 are bound to one another, we require

\begin{equation}
V_{r}\leq\bigg(\frac{2GM_{tot}}{R_{p}}\bigg)^{\frac{1}{2}}(\cos\alpha)^{\frac{1}{2}}\sin\alpha
\label{eq:vr}
\end{equation}

\noindent where $V_{r} = V \sin\alpha$, $R_{p} = R \cos\alpha$, $M_{tot}$ is the combined mass of the two bodies, and $R$ and $V$ are true (3D) positional and velocity separation between the two objects. $V_{r}$ is line-of-sight relative velocity between the two bodies and $R_{p}$ is the projected separation, and $\alpha$ is the projection angle between the plane of the sky and the line that joins the centers of the two objects \citep{BGH,Gre84,CGB,BFK}. This model assumes radial orbits for the clumps, which are assumed to start their evolution at time $t_0=0$ with separation  $R_0=0$, and are moving apart or coming together for the first time in their history, i.e. we are assuming that we are seeing the cluster prior to merging. The only quantity to be determined in Equation \ref{eq:vr} is $\alpha$. We must probe the $V_r-\alpha$ space to define the probability that the system is gravitationally bound for a given projection angle. That probability can be calculate from

\begin{equation}
{\rm P_{bound}} = \int_{\alpha_1}^{\alpha_2} \cos{\alpha}\;d\alpha
\end{equation}

\begin{figure}
\centering
\includegraphics[width=94mm]{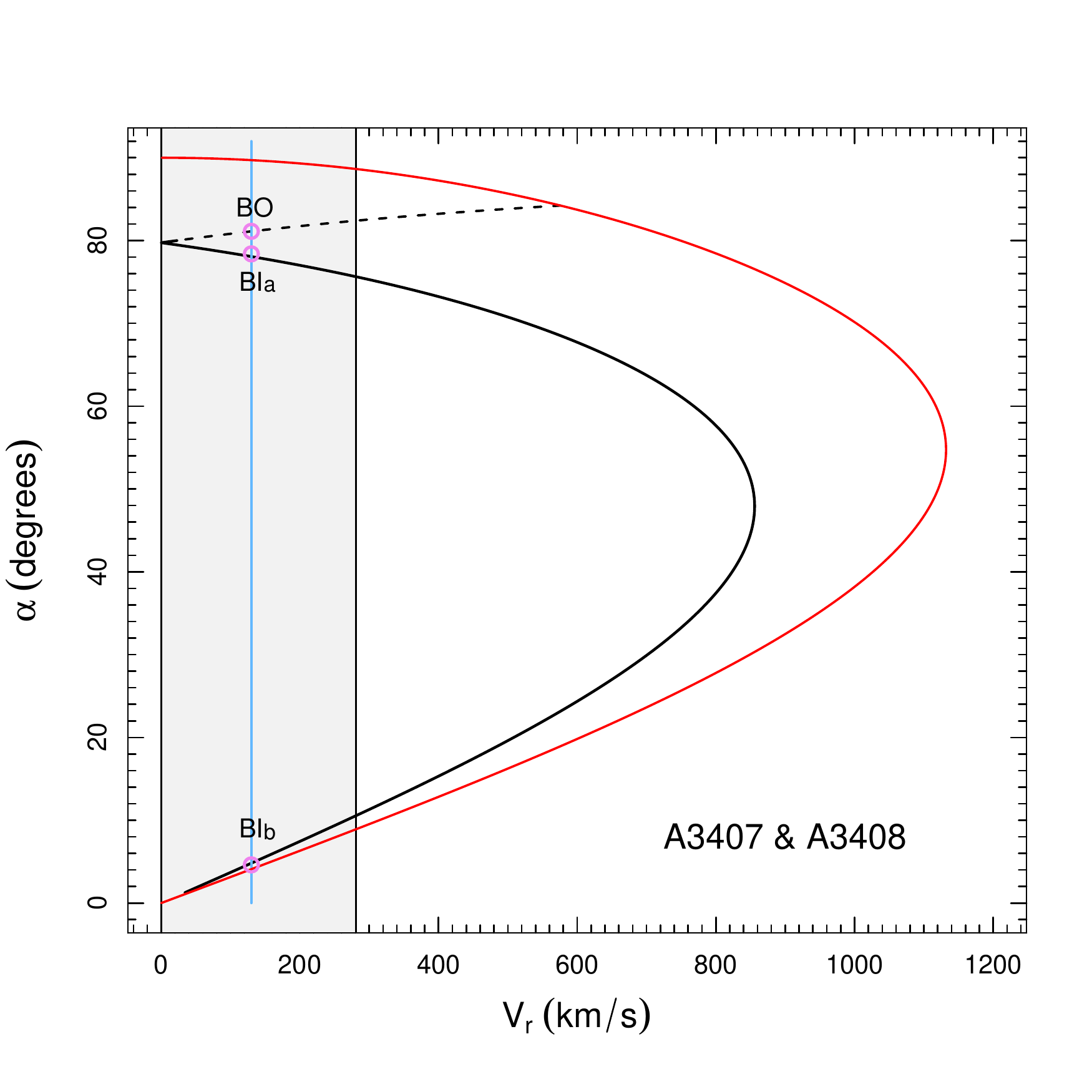}
\caption{Orbits in the two-body model as a function of $\alpha$, the projection angle of A3407 \& A3408, and $V_r$, the radial velocity. The vertical blue line represents the relative radial velocity $V_r=130\pm 151~{\rm km\;s^{-1}}$, with the shaded area indicating the respective uncertainty. The solid curve depicts the bound-incoming solution while the dashed one depicts the bound-outgoing solution. The red line separates the bound and unbound regions according to the Newtonian criterion.
}
\label{fig10}
\end{figure}

\noindent \citep{GDR}. Using the parameters previously found, $R_p=1.68~h^{-1}$ Mpc, $V_r=130~{\rm km\;s^{-1}}$, $M_{tot}=4.53\;h^{-1}\times 10^{14}{\rm M}_\odot$, and taking 9.03 $h^{-1}$ Gyr, as the age of the universe at $z\approx 0.042$, we can solve the two-body problem. The solutions are shown in Figure \ref{fig10}, where the dashed line depicts the bound-outgoing solution (BO) and the solid black line depicts the bound-incoming solution (BI). There are two solutions in the BI case (${\rm BI_a}$ and ${\rm BI_b}$) due to the ambiguity in the projection angle $\alpha$. All the solutions are defined by the vertical blue line corresponding to the relative velocity with the shaded area associated to the error $\pm 151~{\rm km\;s^{-1}}$. The red line in Figure \ref{fig10} separates the bound and unbound regions according to the Newtonian criterion. The general result is that the A3407 and A3408 are likely to be bound at the 84\% level.

The two-body model has three solutions: two collapsing or ingoing and one expanding or outgoing. Considering the angle of parametrization $\chi$ obtained from the spherical collapse model (between $0 < \chi < 2\pi$), we can describe the temporal evolution and the different relative positions of the binary system \citep{Pee93}. These solutions allow us to estimate the time scale for the system to reach the maximum expansion. Using the equations of motion:

\begin{center}
\begin{equation}
t=\bigg(\frac{R_m^3}{8GM}\bigg)^{1/2}(\chi-\sin{\chi}),
\label{eq:t}
\end{equation}
\end{center}

\begin{equation}
R={R_m\over 2} (1 - \cos{\chi}),
\end{equation}

\begin{equation}
V=\left({2GM\over R_m}\right)^{1/2} {\sin{\chi}\over (1 - \cos{\chi})},
\end{equation}

\noindent where $R$ is the separation at time $t$, $R_m$ is the separation at the maximum expansion, and V is the relative velocity. We can solve this system of equations for $V_r$ and $\alpha$ using equation (6) from \citet{Gre84}:

\begin{equation}
\tan{\alpha}=\left({tV_r\over R_p}\right){(\cos{\chi} -1)^2\over \sin{\chi}(\chi - \sin{\chi})}.
\end{equation}

\noindent In Table \ref{tab4}, we present the parameters of the two-body model solutions. In solution ${\rm {BO}}$, the system is initially 4.45 $h^{-1}$ Mpc apart, and would still take $\sim 0.76\;h^{-1}$ Gyr to reach the maximum expansion. In solution ${\rm {BI_a}}$, the system is 5.83 $h^{-1}$ Mpc apart with a low colliding velocity of $135.68$ ${\rm km\;s^{-1}}$. In these two first solutions, the cluster cores will cross each other after a long time ($\geq$ 10 Gyr). The only solution consistent with a close encounter is the ${\rm {BI_b}}$ solution, where clusters are close together with a high colliding velocity of $1080.85$ ${\rm km\;s^{-1}}$, and the cluster cores will cross in less than $\sim 1\;h^{-1}$ Gyr. Thus, while the ${\rm {BO}}$ and ${\rm {BI_a}}$ solutions do not predict strong interactions between A3407 and A3408, the ${\rm {BI_b}}$ solution allows a more intense dynamics for this pair. This latter possibility is examined in section 6.

\begin{table}
\begin{center}
{\small
\caption{Solutions of the two-body model for A3407 \& A3408.}
\label{tab4}
\begin{tabular}{|c c  c  c c c |}
\hline
Sol. & $\alpha $ & $V$ & $R$ & $R_m$ & t \\
\hline
 ~~ & ($\circ$)  & (${\rm km~s^{-1}}$) & ($h^{-1}\;$Mpc) & ($h^{-1}\;$Mpc) & $h^{-1}\;$Gyr \\
\hline
BO & 73.25 & ~~140.83  & 4.45 &  4.56 & 5.38\\
${\rm BI_a}$ & 67.85 & $-$135.68 & 5.83  & 6.00 & 3.56\\
${\rm BI_b}$ & 6.67 &  $-$1080.85 & 1.69 & 3.43 & 8.23\\
\hline
\end{tabular}
}
\end{center}
\label{tb4}
\end{table}

\subsection{Caveats about the method}

Before proceeding, we have to recognise some weaknesses of the two-body model. First of all, we should keep in mind that the two-body model does not consider the angular moment of the system, which is unlikely to be zero, since we have identified a velocity assymetry around the principal axis of the galaxy projected distribution, suggesting a rotating object. The reason this could be important to any dynamical analysis is that the corrected cluster mass can be reduced by $\sim$20\%-30\%, on average, with respect to that uncorrected for rotation, as shown in \citet{MP16}. An additional weakness of the two-body model is not considering the distribution of matter inside each cluster. As clusters merge, their halos overlap and dark matter constraints should be taken into account, as shown in \citet{Nu08}. Finally, we also ignored the gravitational interaction of the infalling matter outside the cluster pair, which could affect the assumption that masses are constant since their formation time \citep{AMc8}. All these points may modify the results presented in section 5.2.

\section{Discussion}

It is not easy to distinguish the most reasonable two-body dynamical solution. Several other factors must be considered to come up with a viable physical scenario. In the present work, the solution that can be best assessed with the available data is ${\rm {BI_b}}$, in which the close proximity of the clusters combined with the high colliding velocity would suggest that the system is about to coalesce, and should already be experiencing some dynamical interactions. If the system was at such evolutionary state, we would expect to see some disturbances in the velocity distributions, which has not been observed in this  work: both clusters have Gaussian velocity distributions (see Section 5.1). In addition, applying the statistical tests described in Section 4, none of them indicates the presence of substructures in A3408 at the 95\% c.l. (with p-values $>$ 0.221), and only
the DS test indicates substructures in A3407 at the 95\% c.l. (p-value=0.016). Hence, these results (taken separately) are not indicating strong dynamical interactions in the pair. 

But cluster dynamics is also related to galaxy evolution. A cluster-cluster interaction may affect galaxy orbits, star formation rates, colors, and morphologies. From this perspective, there are some important differences between A3407 and A3408. A3407 has a higher fraction of emission line galaxies, $\sim 30\%$, in comparison to A3408, $\sim 8\%$. Also, galaxies in A3408 are redder [$\langle B-R \rangle = 1.83 \pm 0.22$] than in A3407 [$\langle B-R \rangle = 1.47 \pm 0.12$], with the KS test indicating color distributions significantly different  between the clusters (p-value$<10^{-4}$). Finally, while the BCG in A3407 has larger magnitude differences with respect to the second and third brightest galaxies [$\Delta m_{12}=0.62$, $\Delta m_{13}=0.75$], in A3408 basically there is no magnitude difference between the three first ranked galaxies [$\Delta m_{12}=0.01$, $\Delta m_{13}=0.01$]. These findings are further examples of the complexity of this pair. A3408 seems to contain more evolved galaxies with significant suppression of the star formation rate. At the same time, this cluster does not have an unquestioned BCG (three galaxies could take the position). This is not in agreement with the central galaxy paradigm, which states that a  BCG, with pronounced luminosity gap, should be at rest at the center of the cluster. The amplitude of the luminosity gap is a function of
the formation epoch, the halo concentration, and the recent infall history of the cluster \citep[see][]{S10}. This could be suggesting a dark matter halo less concentrated or disturbed in A3408. Disturbances in DM correspond well to what we expect from the ${\rm {BI_b}}$ solution, and would probably leave traces in the X-ray emission.

\begin{figure}
\centering
\includegraphics[width=84mm]{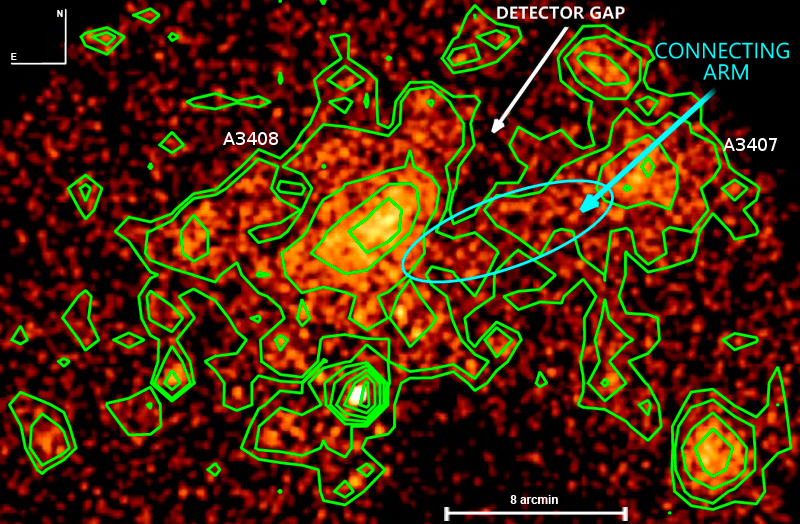}
\caption{X-ray emission around A3407 and A3408 with isophotes in green. A possible ``arm" connecting the clusters is highlighted. ROSAT image. }
\label{fig11}
\end{figure}

Taking a closer look at the X-ray distribution around A3407 and A3408 we note a complex and patchy X-ray morphology extended in the SE-NW direction of this field -- Fi\-gu\-re \ref{fig11}. In A3407, the ROSAT image suggests the presence of substructures, and the cluster core shows an elongation roughly in the same direction as that of A3408. In A3408, the X-ray diffuse emission has an elongation SE-NW (not towards A3407, but with similar elongation direction of its core). We also should note in this figure a weak bridge or ``arm" leaving A3407 and going towards A3408. Since this arm does not seem to be the result of point sources, it may indicate the existence of a physical connection between the clusters. These X-ray features are compelling indicators of dynamical interactions between A3407 and A3408, and suggest the ${\rm BI_b}$ solution as a viable model for the pair. Indeed, the X-ray emission around this field could be also consistent with a merger having happened in the SE-NW direction, which is approximately coincident with the principal axis of the galaxy projected distribution (see Section 3.4). If that is the case, we may be witnessing a major post merging event. Indeed, from the X-ray temperature, \citet{KHH} find $\sigma= 674\pm 23$ ${\rm km\;s^{-1}}$, significantly higher than the velocity dispersion we found from galaxies, $\sigma= 573_{-59}^{+82}$ ${\rm km\;s^{-1}}$. This could mean that the gas temperature may be enhanced by previous-ongoing shock heating due to a merger. Running the two-body model for $2\pi < \chi < 4\pi$ we find two expanding (outgoing) solutions with low (174 ${\rm km\;s^{-1}}$) and high velocities (1135 ${\rm km\;s^{-1}}$), and probability of being a gravitationally bound system of $\sim$ 8\%, with cores having crossed each other $\sim 1.65\;h^{-1}$ Gyr ago. 

\begin{figure}
\centering
\includegraphics[width=90mm]{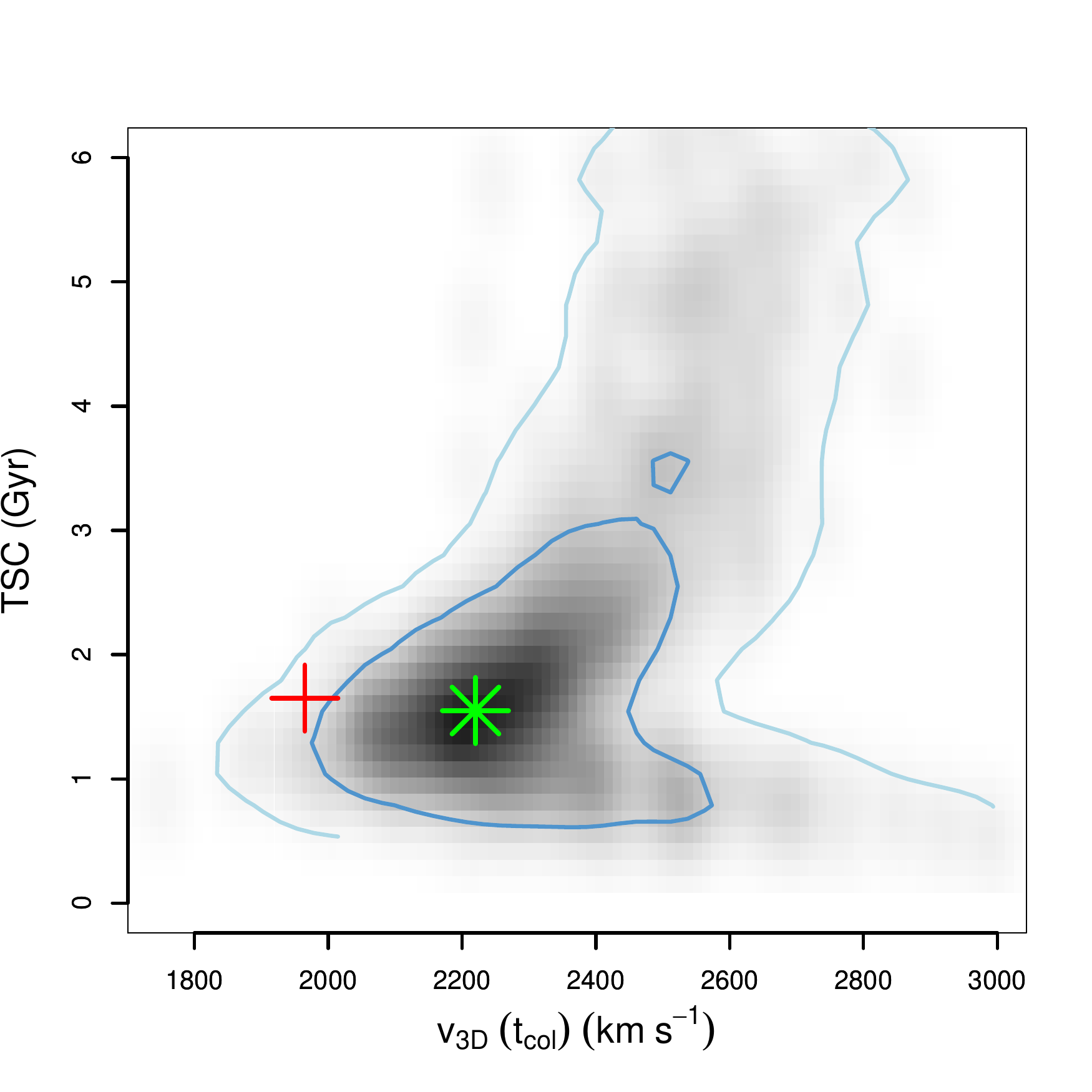}
\caption{The posterior PDF of the A3407 \& A3408 pair for the TSC and ${\rm v_{3D}(t_{col})}$ parameters is
shown in grayscale, with dark and light blue contours representing 68\% and 95\% confidence levels, respectively. The red cross indicates the approximate location of the traditional two-body model
solution. The green star indicates the most probable solution of the MCMAC code for 10,000 realizations.
}
\label{fig12}
\end{figure}

The post-merger scenario can be further explored with the Dawson's method -- the Monte Carlo Merger Analysis Code (MCMAC).\footnote{A Python code openly available at git://github.com/MCTwo/MCMAC.git.} The method takes observed priors on each cluster's mass, radial velocity, and projected separation, draws randomly from those priors, and uses them in a analytic model to get posterior PDF's for merger dynamic properties of interest \citep[see][]{DA13}. This method can obtain a valid solution near the collision state, fully estimate the covariance matrix for the merger parameters, and it is in better than 10\% agreement with N-body simulations of dissociative mergers. A further advantage of the MCMAC method is that the potential gravitational energy of the pair is approximated by two truncated NFW halos, a more realistic setup for interacting clusters \citep[][]{DA13}. Performing the analysis for A3407 \& A3408 with 10,000 Monte Carlo realizations, we find the posterior distribution of the time-since-collision (TSC) and the three-dimensional relative velocity  ${\rm v_{3D}(t_{col})}$ parameters. In Figure 12 we see the posterior PDF of the A3407 \& A3408 pair. Note that the distribution encompasses the two-body model high velocity solution (the red cross). Here, we assume velocity isotropy and transform the line-of-sight relative velocity to the three-dimensional one at the collision time just by multiplying a $\sqrt{3}$ factor. This allows a direct comparison with the MCMAC results. Note in Figure \ref{fig12} that the most probable solution of the MCMAC method has a higher velocity at the collision time (2220 ${\rm km\;s^{-1}}$) and a similar time-since-collision (1.55 $h^{-1}$ Gyr) (the green star). Also, note that a low velocity solution seems to be discarded by the MCMAC method. This result reinforces the possibility of a high velocity post-merger solution for A3407 \& A3408.

\section{Conclusion}

We performed a dynamical study of the galaxy cluster pair A3407 \& A3408 based on a spectroscopic survey obtained with the 4 meter Blanco telescope at the CTIO, plus 6dF data, and also considering X-ray data from the ROSAT All-Sky-Survey. Our main goal was to probe the galaxy dynamics in this field and verify if the sample constitutes a single galaxy system or corresponds to an ongoing merging process. The currently pair description considerably enhances the knowledge about this field, which were previously restricted to the 1990s analyses.

As a central result, the KMM analysis and the X-ray distribution do not support the supposition made by \citet{GCF} that A3407 and A3408 may form a single system. Despite the very regular configuration in the velocity space, our data is consistent with two distinct objects. At the same time, redder galaxy colours and the absence of a unique BCG in A3408 indicates this system could be disturbed, in agreement with the complex X-ray distribution in this field. Signs of dynamical interactions suggest that the systems may be consistent with the fast collapsing solution of the two-body model, configuring a pre-merger scenario. In this scenario the cluster cores will cross each other in less than $\sim 1 h^{-1}$Gyr. On the other hand, the gas temperature and galaxy evolution indicators in A3408 may be suggesting a post-merger scenario, with cores having crossed each other $\sim 1.65 h^{-1}$Gyr ago. This result is reinforced by the MCMAC analysis which provides
the most probable solution (at the $\sim$82\% level) for the post-merger picture with TSC=1.55 $h^{-1}$Gyr and ${\rm v_{3D}(t_{col})=2200~km\;s^{-1}}$. Hence, that possibility can not be rejected in the present work. Further observations and N-body numerical simulations are required to confirm or refute each scenario discussed in this paper.

\section*{Acknowledgments}
We thank the referee for relevant suggestions.
We are grateful to E. Cypriano, R. Monteiro and G. Lima Neto for useful suggestions and discussions. We also thank the financial support of FAPESB under grant 0239/2010. ALBR and HP would like to thank the financial support from the project  Casadinho PROCAD - CNPq/CAPES number 552236/2011-0. ALBR also thanks the support of CNPq, grant 309255/2013-9. MT acknowledges financial support from FAPESP (process 2012/05142-5) and CNPq (process 204870/2014-3).  Finally, we would like to thank the CTIO staff.

\clearpage

\appendix

\onecolumn

\section{Tables I \& II}

\vspace{0.5cm}

Tables 1 and 2 present the properties obtained with the Rvsao and o Xcsao tasks, and the magnitudes, listed in column 3, obtained from Sextractor aperture magnitude with an aperture of 3.2\arcsec, for both cluster.

\begin{longtable}{c c c c c c}
 \caption{ \label{Tab_A3407} Abell 3407} \\
\hline\hline

  R.A.     & Decl.        &       & $V$           &       &          \\
  (J2000)  & (J2000)      & $m_R$ & (km s$^{-1}$) & $R_{quality}$   & Emission Lines \\
  (1)      & (2)          & (3)   & (4)           & (5)   & (6)	    \\

\hline
\endfirsthead
\caption{continued.}\\
\hline\hline

  R.A.     & Decl.        &       & $V$           &       &          \\
  (J2000)  & (J2000)      & $m_R$ & (km s$^{-1}$) & $R_{quality}$   & Emission Lines \\
  (1)      & (2)          & (3)   & (4)           & (5)   & (6)	    \\

\hline
\endhead
\hline

\endfoot
 7:03:28.36 & -49:10:49.3  & 17.90 &  49440 $\pm$  70  	   &  4.90 & ...	    \\ 
7:03:37.58 & -49:03:31.8  & 14.40 &  12865 $\pm$  28  	   & 14.01 & ...	    \\ 
7:03:39.11 & -48:59:29.9  & 17.01 &  13756 $\pm$  28  	   & 14.06 & ...	    \\ 
7:03:42.36 & -49:01:46.3  & 16.99 &  12926 $\pm$  33  	   & 10.60 & ...	    \\ 
7:03:44.49 & -48:59:08.5  & 17.34 &  14551 $\pm$  38  	   &  7.45 & \small{H$\beta$, H$\alpha$, NII}  \\ 
7:03:47.94 & -48:58:25.8  & 18.21 &  12597 $\pm$  39  	   &  3.51 & \small{OII, H$\beta$, H$\alpha$, NII}\\
7:03:48.57 & -49:08:24.6  & 17.05 &  12804 $\pm$  29  	   &  9.28 & ...	    \\ 
7:03:49.09 & -49:18:49.0  & 18.27 &  12215 $\pm$  85  	   &  3.30 & \small{H$\beta$, OIII, OIII, H$\alpha$, NII, SI}   \\
7:03:52.41 & -49:09:55.1  & 18.75 &  49787 $\pm$  55  	   &  5.71 & ...	    \\
7:03:53.70 & -49:06:54.3  & 17.23 &  12615 $\pm$  38  	   &  4.67 & ...	    \\  
7:03:55.17 & -49:03:41.6  & 18.39 &  13620 $\pm$  39  	   &  3.46 & \small{H$\beta$, H$\alpha$, NII}	\\
7:03:58.00 & -49:04:50.0  & 14.23 &  12885 $\pm$  42  	   & 17.74 & ...	    \\ 
7:03:59.45 & -49:05:12.9  & 14.70 &  12621 $\pm$  26  	   &  7.32 & \small{OI, H$\alpha$}   \\ 
 7:04:03.92 & -49:05:59.8  & 16.53 &  13304 $\pm$  22  	   & 18.20 & ...	    \\
 7:04:04.34 & -49:20:45.5  & 17.25 &  13215 $\pm$  37  	   &  5.49 & \small{OI}     \\ 
 7:04:05.96 & -48:54:31.4  & 18.27 &  56185 $\pm$  53  	   &  3.61 & ...	    \\ 
 7:04:08.56 & -49:03:32.3  & 18.22 &  49744 $\pm$  36  	   &  5.71 & \small{OIII, OIII}   \\ 
 7:04:09.02 & -49:10:35.3  & 17.17 &  11623 $\pm$  29  	   &  9.73 & ...	    \\ 
 7:04:11.68 & -49:13:17.4  & 17.78 &  49463 $\pm$  34  	   &  7.40 & ...	    \\ 
 7:04:23.14 & -49:14:01.2  & 18.61 &  49454 $\pm$  88  	   &  3.90 & ...	    \\ 
 7:04:24.92 & -49:20:47.1  & 16.83 &  	-1 $\pm$  43  	   &  4.55 & \small{OII, OI}\\ 
 7:04:25.97 & -48:45:43.3  & 16.38 &  13858 $\pm$  37  	   &  9.68 & \small{OII, OIII}   \\ 
 7:04:27.02 & -49:06:49.1  & 16.77 &  12857 $\pm$  28  	   & 11.65 & \small{OII, OI}\\
 7:04:30.11 & -49:10:54.1  & 17.36 &  11613 $\pm$  26  	   &  9.23 & ...	    \\
 7:04:30.31 & -48:47:03.2  & 15.78 &  13138 $\pm$  35  	   & 13.26 & \small{OI, SII} \\
 7:04:32.50 & -49:03:12.6  & 18.51 &  12589 $\pm$  49  	   &  3.61 & ...	    \\ 
 7:04:34.43 & -48:59:35.6  & 17.70 &   76   $\pm$  39  	   &  4.45 & ...	    \\ 
 7:04:36.80 & -49:04:31.9  & 17.15 &  11951 $\pm$  45  	   & 11.54 & ...	    \\ 
 7:04:37.41 & -49:01:29.8  & 18.43 &  11966 $\pm$  37  	   &  4.10 & ...	    \\ 
 7:04:38.07 & -48:58:57.1  & 16.55 &  12820 $\pm$  34  	   & 14.70 & ...	    \\ 
 7:04:39.08 & -48:49:48.6  & 18.17 &  13481 $\pm$  58  	   &  5.18 & \small{OI}     \\ 
 7:04:42.24 & -49:06:51.1  & 16.30 &  12931 $\pm$  24  	   & 19.67 & ...	    \\ 
 7:04:43.40 & -48:56:43.9  & 18.34 &  12957 $\pm$  93  	   &  3.60 & ...	    \\ 
 7:04:43.93 & -49:02:46.7  & 16.47 &  13175 $\pm$  29  	   & 12.93 & ...	    \\
 7:04:46.94 & -49:09:50.7  & 16.30 &  11591 $\pm$  33  	   & 14.55 & ...	    \\ 
 7:04:47.05 & -48:45:36.4  & 17.38 &  12823 $\pm$  56  	   &  5.37 & \small{OII}    \\ 
 7:04:52.15 & -49:05:51.3  & 17.10 &  12105 $\pm$  23  	   & 11.82 & ...	    \\ 
 7:04:56.44 & -49:06:18.8  & 17.61 &  10733 $\pm$  56  	   &  5.10 & ...	    \\ 
 7:04:58.15 & -49:02:31.6  & 17.38 &  32285 $\pm$  33  	   &  8.29 & \small{OIII}   \\ 
 7:04:58.58 & -49:21:37.1  & 16.40 &  13558 $\pm$  33  	   & 12.92 & \small{OI}     \\ 
 7:05:04.76 & -49:00:18.5  & 17.43 &  11867 $\pm$  33  	   &  9.29 & ...	    \\ 
 7:05:05.77 & -48:51:08.1  & 17.24 &  12439 $\pm$  36  	   &  2.99 & \small{OII, H$\beta$, OIII, OIII, H$\alpha$, NII}		    \\ 
 7:05:03.18 & -49:06:01.5  & 18.24 &  11294 $\pm$  36  	   &  5.54 & ...	    \\ 
 7:05:09.45 & -49:02:51.7  & 15.26 &  11289 $\pm$  33  	   &  9.17 & ...	    \\ 
 7:05:10.25 & -49:00:17.6  & 17.68 &  12229 $\pm$  28  	   &  7.27 & ...	    \\ 
 7:05:11.10 & -48:53:35.2  & 17.46 &  21859 $\pm$  55  	   &  5.84 & \small{OII, H$\beta$, OIII, OIII} \\
 7:05:14.41 & -49:17:15.6  & 17.78 &  32419 $\pm$  48  	   &  6.40 & \small{OIII}   \\
 7:05:13.18 & -49:13:19.6  & 16.66 &  11419 $\pm$  26  	   & 11.16 & ...	    \\ 
 7:05:16.07 & -49:16:39.9  & 17.33 &  13193 $\pm$  42  	   &  5.60 & \small{OII, OIII, OIII, OI, H$\alpha$}\\ 
 7:05:20.99 & -49:07:34.6  & 17.98 &  11360 $\pm$  53  	   &  4.11 & ...	    \\ 
 7:05:23.38 & -49:00:12.5  & 17.69 &  12602 $\pm$  22  	   &  8.97 & ...	    \\ 
 7:05:29.04 & -49:07:40.8  & 16.08 &  12113 $\pm$  30  	   & 13.13 & ...	    \\ 
 7:05:30.99 & -48:58:19.0  & 18.56 &  21833 $\pm$  70  	   &  3.10 & \small{OII, H$\beta$, OIII, OIII} \\ 
 7:05:35.55 & -49:04:14.0  & 17.51 &  11718 $\pm$  33  	   &  8.11 & ...	    \\ 
 7:05:36.69 & -49:19:57.9  & 17.23 &  13538 $\pm$  39  	   &  5.07 & ...	    \\    
 7:05:37.22 & -49:17:17.2  & 17.86 &  41646 $\pm$  46  	   &  3.96 & ...	    \\ 
 7:05:39.59 & -48:45:15.6  & 16.70 &  12693 $\pm$  43  	   &  3.39 & \small{H$\beta$, OI, H$\alpha$, NII}  \\ 
 7:05:47.66 & -48:57:40.2  & 14.63 &  12567 $\pm$  34  	   & 10.03 & \small{OII}    \\ 
 7:05:49.42 & -49:02:52.2  & 17.56 &  11755 $\pm$  42  	   &  7.08 & ...	    \\ 
 7:05:55.46 & -48:47:10.8  & 17.58 &  12775 $\pm$ 116 	   &  3.30 & \small{OIII}   \\  
 7:05:58.44 & -49:15:00.5  & 17.10 &  11959 $\pm$  41  	   &  6.92 & ...	    \\ 
 7:06:01.47 & -49:00:59.6  & 16.81 &  11844 $\pm$  41  	   &  6.71 & ...	    \\
 7:06:11.81 & -49:08:11.2  & 16.57 &   20   $\pm$  41  	   &  5.86 & ...	    \\ 
 7:06:14.75 & -49:02:51.8  & 16.26 &  11694 $\pm$  28  	   & 11.61 & \small{OII, S1}\\
 7:06:19.08 & -48:49:19.9  & 16.62 &  12320 $\pm$  54  	   &  5.53 & \small{OI}     \\ 
 7:06:33.93 & -49:04:25.4  & 17.45 &  12858 $\pm$  46  	   &  5.07 & ...	    \\ 
 7:06:50.94 & -49:04:09.3  & 17.26 &  11897 $\pm$  39  	   &  4.16 & \small{H$\alpha$, NII, SI}  \\ 
\hline\hline 
\end{longtable}

\clearpage

\begin{table}
\caption{\label{Tab_A3408} Abell 3408}
 \centering 
 \small
 \begin{tabular}[!h]{c c c c c c}
 \hline\hline
 \rule{0cm}{0.45cm}
  R.A.      & Decl.        &       & $V$             &       &                \\
  (J2000)   & (J2000)      & $m_R$ & (km s$^{-1}$)   & $R_{quality}$   & Emission Lines \\
  (1)       & (2)          & (3)   & (4)             & (5)   & (6)	      \\
 \hline
  7:07:01.99 & -49:19:49.9  & 16.48  &   13207  $\pm$  29	     &  8.36 & ...              \\ 
 7:07:18.86 & -49:25:58.3  & 18.89  &   37652  $\pm$  61      &  2.04 & \small{OII, OIII, H$\beta$} \\
 7:07:20.40 & -49:13:39.2  & 18.87  &   11584  $\pm$ 115	     &  2.60 & ...		 \\  
 7:07:24.17 & -49:21:53.1  & 17.99  &   13193  $\pm$  65      &  2.00 & \small{OI, OII, OIII, H$\alpha$, H$\beta$, NII} \\
 7:07:25.88 & -49:12:50.8  & 15.84  &   13014  $\pm$  44	     &  6.32 & ...		 \\
 7:07:26.27 & -49:12:13.5  & 18.21  &   13009  $\pm$  35	     &  5.31 & ... \\
 7:07:33.05 & -49:21:45.5  & 16.21  &   13226  $\pm$  29	     &  8.33 & \small{OI,OII}  \\ 
 7:07:34.36 & -49:24:51.1  & 16.72  &   12649  $\pm$  33	     &  5.74 & ...		\\  
 7:07:36.08 & -49:10:29.7  & 17.58  &   28150  $\pm$  32	     &  6.42 & ...		 \\ 
 7:07:38.37 & -49:01:35.6  & 18.07  &   12728  $\pm$  62      &  2.65 & \small{OI, OII, OIII, H$\alpha$, H$\beta$, NII} \\
 7:07:39.38 & -49:07:02.7  & 15.37  &   13147  $\pm$  24	     &  9.84 & ...		\\ 
 7:07:39.50 & -49:14:12.8  & 17.80  &   12127  $\pm$  28	     &  6.82 & ...		\\ 
 7:07:45.19 & -49:24:59.6   & 17.14 &   12512  $\pm$  36	     &  6.21 & ...              \\ 
 7:07:52.22 & -49:07:25.3  & 17.40  &   12753  $\pm$  30	     &  8.33 & ... \\ 
 7:07:59.17 & -49:09:56.6  & 16.82  &   12110  $\pm$  26	     & 11.50 & ...		\\ 
 7:07:59.46 & -49:01:36.1  & 17.09  &   13169  $\pm$  44	     &  4.40 & ...		\\ 
 7:07:59.59 & -49:16:54.9  & 17.06  &   12697  $\pm$  37	     &  9.00 & ...            \\
 7:08:06.10 & -49:24:21.5  & 18.28  &   36539  $\pm$  61	     &  3.01 & ... \\ 
 7:08:06.84 & -49:15:39.1  & 18.65  &   12660  $\pm$  83	     &  2.60 & \small{OIII, H$\alpha$} \\ 
 7:08:08.82 & -49:10:25.4  & 14.80  &   13027  $\pm$  29	     &  6.00 & ...		\\ 
 7:08:10.60 & -48:58:08.8  & 18.86  &   11935  $\pm$  75      &  1.80 & \small{OI, OII, H$\alpha$, NII} \\
 7:08:11.06 & -49:14:17.5  & 18.62  &   42322  $\pm$  48	     &  3.90 & ...		\\
 7:08:11.44 & -49:09:53.1  & 13.50  &   12653  $\pm$  24	     & 12.98 & ...		 \\ 
 7:08:14.25 & -49:08:09.3  & 15.98  &   11901  $\pm$  40	     & 13.85 & ...		\\ 
 7:08:16.75 & -49:11:44.9  & 16.18  &   12934  $\pm$  25	     &  9.47 & ...             \\ 
 7:08:21.71 & -49:07:07.5  & 15.97  &   11867  $\pm$  26	     &  9.25 & \small{OII}    \\
 7:08:24.25 & -49:11:49.1  & 17.37  &   13524  $\pm$  23	     & 11.10 & ...		\\
 7:08:30.01 & -49:20:07.6  & 16.72  &   13281  $\pm$  32	     &  8.11 & ...		 \\  
 7:08:31.80 & -49:06:46.3  & 16.54  &   10772  $\pm$  30	     & 10.11 & ... \\
 7:08:35.36 & -49:12:59.6  & 16.70  &   11458  $\pm$  40	     &  3.74 & ...		 \\ 
 7:08:37.34 & -49:08:19.7  & 18.23  &   12793  $\pm$  65      &  2.07 & \small{OI, OII, OIII, H$\alpha$, H$\beta$, NII} \\
 7:08:45.66 & -49:03:02.9  & 17.33  &   12168  $\pm$  46	     &  2.79 & ...		\\ 
 7:08:47.16 & -49:13:05.6  & 17.57  &   13610  $\pm$  29	     &  4.88 & ... \\ 
 7:08:47.19 & -49:17:29.5  & 17.77  &   21830  $\pm$  55	     &  3.85 & ...             \\ 
 7:08:47.42 & -49:00:14.4  & 18.86  &   46739  $\pm$  76      &  1.98 & \small{OII, OIII, H$\beta$} \\
 7:08:55.32 & -49:16:18.0  & 16.28  &   12764  $\pm$  32	     &  7.32 & ...		\\  
 7:08:56.03 & -49:18:07.0  & 16.92  &   12417  $\pm$  32	     &  5.68 & ... \\  
 7:10:18.03 & -49:07:43.7  & 15.50  &   13391  $\pm$  42	     &  3.29 & ... \\

 \hline\hline 
 
\end{tabular}
 
\end{table}



\label{lastpage}


\begin{thebibliography}{99}


\bibitem[\protect\citeauthoryear{Abell et al.}{1989}]{Abe89} Abell G.O., Corwin, H. G., Olowin, R. P., 1989, ApJS, 70, 1

\bibitem[\protect\citeauthoryear{Allen et al.}{2011}]{A11} Allen S.~W., Evrard, A.~E., Mantz, A.~B. 2011, ARA\&A, 49, 409	

\bibitem[\protect\citeauthoryear{Angus \& McGaugh}{2008}]{AMc8} Angus G. W., McGaugh, S. S. 2008, MNRAS, 383, 417A

\bibitem[\protect\citeauthoryear{Araya-Melo}{2009}]{AM9} Araya-Melo P.~A., Reisenegger, A., Meza, A., 2009, MNRAS, 399, 97

\bibitem[\protect\citeauthoryear{Ashman, Bird \& Zepf}{1994}]{ABZ} Ashman K. M., Bird, C. M., Zepf, S. E., 1994, AJ, 108, 2348  

\bibitem[\protect\citeauthoryear{Barden \& Ingerson}{1998}]{Bar98} Barden S.C. \& Ingerson, T.E., 1998, ASP Conf. 152, eds. S. Arribas, E. Mediavilla, and F. Watson (ASP), 60

\bibitem[\protect\citeauthoryear{Barne \& Williamss}{2012}]{BW12} Barnes E. I. \& Williams, L. L. R., 2012, ApJ, 748, 144

\bibitem[\protect\citeauthoryear{Barrena, Biviano \& Ramella }{2002}]{Ba2} Barrena R., Biviano, A., Ramella, M., et al. 2002, A\&A, 386, 816 

\bibitem[\protect\citeauthoryear{Beers, Geller \& Huchra}{1982}]{BGH} Beers T. C., Geller, M. J., Huchra, J. P., 1982, ApJ, 257, 23 

\bibitem[\protect\citeauthoryear{Beers, Flynn \& Gebhart}{1990}]{Bee90} Beers T. C., Flynn, K., Gebhardt, K., 1990, AJ, 100, 32

\bibitem[\protect\citeauthoryear{Beers, Gebhart \& forman}{1991}]{BGF} Beers T, C., Gebhardt, K., Forman, W., 1991, AJ, 102, 1581

\bibitem[\protect\citeauthoryear{Beraldo e Silva, Lima \& Sodr\'e }{2013}]{BLS} Beraldo e Silva L., Lima, M., Sodr\'e, L., 2013, MNRAS, 436, 2616

\bibitem[\protect\citeauthoryear{Bertin \& Arnouts}{1996}]{Ber96} Bertin E. \& Arnouts, S., 1996, A\&AS, 317, 393 

\bibitem[\protect\citeauthoryear{Bird \& Beers}{1993}]{BB} Bird C.~M. \& Beers, T.~C.,1993, AJ, 105, 1596

\bibitem[\protect\citeauthoryear{Biviano, Murante \& Borgani}{2006}]{BMB} Biviano A., Murante, G., Borgani, S., et al. 2006, A\&A, 456, 23 

\bibitem[\protect\citeauthoryear{Brough, Forbes \& Kilborn}{2006}]{BFK} Brough S., Forbes, D. A., Kilborn, V. A., 2006, MNRAS, 369, 1351

\bibitem[\protect\citeauthoryear{Campusano, Kneib \& Hardy}{1998}]{CKH} Campusano L. E., Kneib, J. P., Hardy, E., 1998, ApJL, 496, L79

\bibitem[\protect\citeauthoryear{Campusano \& Hardy}{1996}]{CH} Campusano L. E. \& Hardy, E., 1996, {\it Astrophysical Applications of Gravitational Lensing}, p.125, Kluwer (eds. Kochanek, C.S. and Hewitt, J. N.)

\bibitem[\protect\citeauthoryear{Carrasco, Mendes de Oliveira \& Infante}{2006}]{Car06} Carrasco E., R., Mendes de Oliveira, C., Infante, L., 2006, AJ, 132, 1796

\bibitem[\protect\citeauthoryear{Carter \& Metcalfe}{1980}]{CM} Carter D. \& Metcalfe, N., 1980, MNRAS, 190, 325

\bibitem[\protect\citeauthoryear{Cattaneo, Mamon \& Warnick}{2011}]{CMW} Cattaneo A., Mamon, G.~A., Warnick, K., 2011, A\&A, 533, A5

\bibitem[\protect\citeauthoryear{Clowe, Bradac \& Gonzalez}{2006}]{CBG} Clowe D., Bradac, M., Gonzalez, A. H., 2006, ApJ, 648, L109
 
\bibitem[\protect\citeauthoryear{Cohn \& White}{2005}]{CW} Cohn J.~D. \& White, M. 2005, APh, 24, 316

\bibitem[\protect\citeauthoryear{Cortese, Gavazzi \& Boselli}{2004}]{CGB} Cortese L., Gavazzi, G., Boselli, A., 2004, A\&A, 425, 429

\bibitem[\protect\citeauthoryear{Croux \& Dehon}{2013}]{CD} Croux C. \& Dehon C., 2013, EE, ohn Wiley \& Sons, Ltd

\bibitem[\protect\citeauthoryear{Cypriano, Sodr\'e \& Campusano}{2001}]{CSC} Cypriano E. S., Sodr\'e, L. Jr., Campusano, L. E., et al. 2001, AJ, 121, 10 

\bibitem[\protect\citeauthoryear{Dawson}{2013}]{DA13} Dawson, W. A. 2013, ApJ, 772, 131

\bibitem[\protect\citeauthoryear{De Lucia, Springel \& White}{2006}]{DSW} De Lucia G., Springel, V., White, S.~D.~M.,2006, MNRAS,  366, 499

\bibitem[\protect\citeauthoryear{den Hartog \& Katgert}{1996}]{DK} den Hartog R. \& Katgert, P., 1996, MNRAS, 279, 349

\bibitem[\protect\citeauthoryear{Dressler \& Shectman} {1988}]{Dre88} Dressler A. \& Shectman S. A., 1988, AJ, 95, 985
 
\bibitem[\protect\citeauthoryear{Ebeling, Voges \& Bohringer}{1996}]{EVB} Ebeling H., Voges, W., Bohringer, H., et al. 1996, MNRAS, 281, 799

\bibitem[\protect\citeauthoryear{Escalera, Biviano \& Girardi}{1994}]{EBG} Escalera E, Biviano, A., Girardi, M., et al. 1994, ApJ, 423, 539 

\bibitem[\protect\citeauthoryear{Fadda, Girardi \& Giuricin}{1996}]{FGG} Fadda D., Girardi, M., Giuricin, G., et al. 1996, ApJ, 473, 670

\bibitem[\protect\citeauthoryear{Fichett \& Webster} {1987}]{Fic87} Fichett M.J. \&  Webster, R., 1987, ApJ, 317, 653

\bibitem[\protect\citeauthoryear{Fraley \& Raftery}{2006}]{FR} Fraley C. \& Raftery, A.E., 2006, Technical Report, 504

\bibitem[\protect\citeauthoryear{Fujita, Koyama \& Tsuru}{1996}]{FKT} Fujita Y., Koyama, K., Tsuru, T., et al. 1996, PASJ, 48, 191

\bibitem[\protect\citeauthoryear{Fujita, Tawa \& Hayashida}{2008}]{FTH} Fujita Y., Tawa, N., Hayashida, K., et al. 2008, PASJ, 60, 343

\bibitem[\protect\citeauthoryear{Galli, Cappi \& Focardi}{1993}]{GCF} Galli M., Cappi, A., Focardi, P., et al. 1993, A\&AS, 101, 259 

\bibitem[\protect\citeauthoryear{Girardi, Demarco \& Rosati}{2005}]{GDR} Girardi M., Demarco, R., Rosati, P., et al. 2005, A\&A, 436, 29 

\bibitem[\protect\citeauthoryear{Girardi, Fadda \& Escalera}{1997}]{GFE} Girardi M., Fadda, D., Escalera, E., et al. 1997, ApJ, 490, 56

\bibitem[\protect\citeauthoryear{ Girardi, Giuricin \& Mardirossian}{1998}]{GGM} Girardi M., Giuricin, G., Mardirossian, F., et al. 1998, ApJ, 505,74

\bibitem[\protect\citeauthoryear{Gonz\'alez-Casado, Mamon \& Salvador-Sole}{1984}]{GMS} Gonz\'alez-Casado G., Mamon, G. A., Salvador-Sole, E., 1984, ApJ, 433, 61

\bibitem[\protect\citeauthoryear{Gregory \& Thompson} {1984}]{Gre84} Gregory S. A. \& Thompson L. A., 1984, ApJ, 286, 422

\bibitem[\protect\citeauthoryear{Hansen, Egli \& Hollenstei}{2005}]{HEH} Hansen S. H., Egli, D., Hollenstein, L., et al. 2005, NewA, 10, 379

\bibitem[\protect\citeauthoryear{Hansen, Moore \& Zemp}{2006}]{HMZ} Hansen S. H., Moore, B., Zemp, M., et al. 2006, JCAP, 01, 014

\bibitem[\protect\citeauthoryear{Hartigan \& Hartigan}{1985}]{HH} Hartigan J. A. \& Hartigan, P. M., 1985, Ann. Statist., 13, 70

\bibitem[\protect\citeauthoryear{Heisler, Tremaine \& Bahcall}{1985}]{HTB} Heisler J., Tremaine, S., Bahcall, J. N., 1985, ApJ, 298, 8

\bibitem[\protect\citeauthoryear{Hjorth \& Williams}{2010}]{HW} Hjorth J. \& Williams, L. L. R., 2010, ApJ, 722, 851

\bibitem[\protect\citeauthoryear{Hou et al.} {2009}]{Hou09} Hou A., Parker L., Harris W., et al. 2009, ApJ, 702, 1199

\bibitem[\protect\citeauthoryear{Holzmann \& Vollmer}{2008}]{HV} Holzmann H. \& Vollmer, S., 2008, Advances in Statistical Analysis, 92, 57

\bibitem[\protect\citeauthoryear{Huchra \& Geller}{1982}]{HG} Huchra J. P. \& Geller, M. J., 1982, ApJ, 257, 423

\bibitem[\protect\citeauthoryear{Hwang \& Lee}{2007}]{HL} Hwang H. S. \& Lee, M. G. 2007, ApJ, 662, 236

\bibitem[\protect\citeauthoryear{Ihaka \& Gentleman}{1996}]{IG} Ihaka R. \& Gentleman, R. 1996, JCGS, 5, 299

\bibitem[\protect\citeauthoryear{Jee et al.}{2005a}]{Jee05a} Jee M. J., White R. L., Ben\'itez N., et al., 2005a, ApJ, 618,46

\bibitem[\protect\citeauthoryear{Jee et al.}{2005b}]{Jee05b} Jee M. J., White R. L., Ford H. C., et al., 2005b, ApJ, 634, 813

\bibitem[\protect\citeauthoryear{Jee et al.}{2007}]{Jee07} Jee M. J., Ford, H. C., Illingworth, G. D., et al., 2007, ApJ, 661, 728

\bibitem[\protect\citeauthoryear{Jones, Read \& Saunders}{2009}]{JRS} Jones D. H.,  Read, M. A., Saunders, W., et al. 2009, MNRAS, 399, 683 

\bibitem[\protect\citeauthoryear{Katayama, Hayashida \& Hashimotodani}{2001}]{KHH} Katayama H., Hayashida, K., Hashimotodani, K., 2001, PASJ, 53, 1133

\bibitem[\protect\citeauthoryear{Kato et al.}{2015}]{KA15} Kato Y., Nakazawa, K., Gu, L., et al. 2015, PASJ, 67, 71

\bibitem[\protect\citeauthoryear{Kravtsov \& Borgani}{2012}]{KB} Kravtsov A.~V. \& Borgani, S., 2012, ARA\&A, 50, 353

\bibitem[\protect\citeauthoryear{Krause, Ribeiro \& Lopes}{2013}]{KRL} Krause M.~O., Ribeiro, A.~L.~B., Lopes, P.~A.~A., 2013, A\&A, 551, A143

\bibitem[\protect\citeauthoryear{Kurtz \& Mink}{1998}]{Kur98} Kurtz M. J. \& Mink D. J., 1998, PASP, 110, 934

\bibitem[\protect\citeauthoryear{Landolt}{1992}]{Lan92} Landolt A. U., 1992, AJ, 104, 340

\bibitem[\protect\citeauthoryear{Lazzati \& Chincarini}{1998}]{LC} Lazzati, D. \& Chincarini, G. 1998, A\&A, 339, 52

\bibitem[\protect\citeauthoryear{Lee}{1979}]{Lee79} Lee K.L., 1979, JASA, 74, 708

\bibitem[\protect\citeauthoryear{Li }{1998}]{LI98} Li L.-X., 1998, GReGr, 30, 3

\bibitem[\protect\citeauthoryear{Liao et al. }{2014}]{LI14} Liao, s., Cheng, D., Chu M-c. , Tang, J. 2014, arXiv:1412.3515

\bibitem[\protect\citeauthoryear{Lynden-Bell}{1967}]{LY67} Lynden-Bell, D., 1967, MNRAS, 136, 101

\bibitem[\protect\citeauthoryear{Lopes et al.}{2009}]{L09} Lopes P. A. A., de Carvalho, R. R., Kohl-Moreira, J. L., et al. 2009, MNRAS, 399, 2201

\bibitem[\protect\citeauthoryear{Mahdavi et al.}{2007}]{MA07} Mahdavi A., Hoekstra, H., Babul, A., et al. 2007, ApJ, 668, 806

\bibitem[\protect\citeauthoryear{Markevitch et al.}{2005}]{MA05} Markevitch M., Govoni, F., Brunetti, G., et al. 2005, ApJ, 627, 733

\bibitem[\protect\citeauthoryear{Merrall \& Henriksen}{2003}]{MH} Merrall T. E. C. \& Henriksen, R. N, 2003, ApJ, 595, 43

\bibitem[\protect\citeauthoryear{Miralda-Escude}{1991}]{MI91} Miralda-Escude J., 1991, ApJ, 370, 1

\bibitem[\protect\citeauthoryear{Manolopoulou \& Plionis}{2016}]{MP16} Manolopoulou M. \& Plionis M., 2016, MNRAS arXiv:160406256M

\bibitem[\protect\citeauthoryear{Naab, Johansson \& Ostriker}{2007}]{NJO} Naab T., Johansson, P.~H., Ostriker, J.~P., 2007, ApJ, 658, 710

\bibitem[\protect\citeauthoryear{Nolthenius \& White}{1987}]{NW} Nolthenius R. \& White, S. D. M., 1987, MNRAS, 225, 5050

\bibitem[\protect\citeauthoryear{Nusser}{2008}]{Nu08} Nusser, A. 2008, MNRAS, 384, 343

\bibitem[\protect\citeauthoryear{Oegerle \& Hill}{1992}]{Oeg92} Oegerle W. R. \& Hill, J. M. 1992, AJ, 104, 6

\bibitem[\protect\citeauthoryear{Ogorodnikov}{1957}]{OG57} Ogorodnikov, K. F., 1957, SvA, 1, 748

\bibitem[\protect\citeauthoryear{Owers et al.}{2009}]{OW09} Owers M. S., Nulsen, P. E. J., Couch, W. J., et al., 2009, ApJ, 692, 702 

\bibitem[\protect\citeauthoryear{Owers, Nulsen \& Couch}{2011}]{ONC} Owers M. S., Nulsen, P. E. J., Couch, W. J., 2011, ApJ, 741, 122

\bibitem[\protect\citeauthoryear{Pawl, Evrard \& Dupke}{2005}]{PED} Pawl A., Evrard, A. E., Dupke, R. A., 2005, ApJ, 631, 773

\bibitem[\protect\citeauthoryear{Peebles}{1993}]{Pee93} Peebles P. J. E., 1993, (Princeton University Press)
  
\bibitem[\protect\citeauthoryear{Pinkney et al. }{1996}]{Pin96} Pinkney J., Roettiger, K, Burns, J. O., et al. 1996, ApJS, 104, 1

\bibitem[\protect\citeauthoryear{Ramella et al.}{2007}]{RA07} Ramella M., Biviano, A., Pisani, A., et al. 2007, A\&A, 470, 39

\bibitem[\protect\citeauthoryear{Ribeiro, Lopes \& Trevisan }{2011}]{RLP} Ribeiro A. L. B., Lopes, P.A.A., Trevisan, M., 2011, MNRAS, 413, L81.

\bibitem[\protect\citeauthoryear{Ribeiro, de Carvalho \& Trevisan}{2013}]{RDT} Ribeiro A. L. B., de Carvalho, R. R., Trevisan, M., et al. 2013, MNRAS, 434, 784

\bibitem[\protect\citeauthoryear{Ricker}{1998}]{RI98} Ricker P. M. 1998, ApJ, 499, 670

\bibitem[\protect\citeauthoryear{Ricker \& Sarazin}{2001}]{RS} Ricker P. M. \& Sarazin, C. L., 2001, ApJ, 561, 621

\bibitem[\protect\citeauthoryear{Smith et al. }{2010}]{S10} Smith, G. P., Khosroshahi, H. G., Dariush, A., et al. 2010, MNRAS, 409, 169

\bibitem[\protect\citeauthoryear{Springel, White \& Jenkins}{2005}]{SWJ} Springel V., White, S.~D.~M., Jenkins, A., 2005, Nature, 435, 629 

\bibitem[\protect\citeauthoryear{Takizawa}{2000}]{TA00} Takizawa M., 2000, ApJ, 532, 183 

\bibitem[\protect\citeauthoryear{Teague, Carter \& Gray}{1990}]{TCG} Teague P. F., Carter, D., Gray, P. M., 1990, ApJS, 72, 715	

\bibitem[\protect\citeauthoryear{Tonry \& Davis}{1979}]{Ton79} Tonry J. \& Davis M. 1979, AJ, 84, 1511

\bibitem[\protect\citeauthoryear{Ueda, Itoh \& Suto}{1993}]{UIS} Ueda H., Itoh, M, Suto, Y., 1993, ApJ, 408, 3

\bibitem[\protect\citeauthoryear{van Dokkum}{2001}]{Van01} van Dokkum, P.G., 2001, PASP, 113, 1420

\bibitem[\protect\citeauthoryear{Voit}{2005}]{V05} Voit, G. M, 2005, RvMP, 77, 207

\bibitem[\protect\citeauthoryear{Yahil \& Vidal}{1977}]{Yah77} Yahil A. \& Vidal N. V., 1977, ApJ, 214, 347

\bibitem[\protect\citeauthoryear{Wainer \& Shacht}{1978}]{WS} Wainer H., \& Shacht, S. 1978, Psichometrika, 43, 203 

\bibitem[\protect\citeauthoryear{Werner et al.}{2008}]{WE08} Werner N., Finoguenov, A., Kaastra, J.~S., et al. 2008, A\&A, 482, L29 

\bibitem[\protect\citeauthoryear{West \& Bothun}{1990}]{WB} West M.J. \& Bothun, G.D., 1990 Astrophys. J. 350, 36.

\bibitem[\protect\citeauthoryear{West, Oemler \& Dekel}{1988}]{Wes88} West M.J., Oemler A., Dekel A., 1988, ApJ, 327, 1

\bibitem[\protect\citeauthoryear{Wojtak et al.}{2008}]{WO08} Wojtak R., Lokas, E. L., Mamon, G. A., et al. 2008, MNRAS, 388, 815

\bibitem[\protect\citeauthoryear{Wing \& Blanton}{2013}]{WB} Wing J. D. \& Blanton, E. L., 2013, ApJ, 767, 102


\end{thebibliography}
\end{document}